\documentclass[twocolumn,superscriptaddress,longbibliography]{revtex4-1}
\usepackage{amsmath,amssymb,bm,mathrsfs,graphicx, braket, times,color}
\usepackage[colorlinks=true,citecolor=blue,linkcolor=blue]{hyperref}
\usepackage{soul}
\usepackage{xcolor}

\usepackage{colortbl}
\usepackage[makeroom]{cancel}
\usepackage{array,multirow,graphicx}
\usepackage{tikz}
\usetikzlibrary{shadings}

\begin{document}

\title{Dirac-vortex topological cavity}

\author{Xiaomei Gao$^\dagger$}
\affiliation{Institute of Physics, Chinese Academy of Sciences/Beijing National Laboratory for Condensed Matter Physics, Beijing 100190, China}
\affiliation{MOE Key Laboratory of Weak-Light Nonlinear Photonics, TEDA Institute of Applied Physics and School of Physics, Nankai University, Tianjin 300457, China}
\author{Lechen Yang$^\dagger$}
\affiliation{Institute of Physics, Chinese Academy of Sciences/Beijing National Laboratory for Condensed Matter Physics, Beijing 100190, China}
\affiliation{School of Physical Sciences, University of Chinese Academy of Sciences, Beijing 100049, China}
\author{Hao Lin}
\affiliation{Institute of Physics, Chinese Academy of Sciences/Beijing National Laboratory for Condensed Matter Physics, Beijing 100190, China}
\affiliation{School of Physical Sciences, University of Chinese Academy of Sciences, Beijing 100049, China}
\author{Lang Zhang}
\affiliation{Institute of Physics, Chinese Academy of Sciences/Beijing National Laboratory for Condensed Matter Physics, Beijing 100190, China}
\affiliation{MOE Key Laboratory of Weak-Light Nonlinear Photonics, TEDA Institute of Applied Physics and School of Physics, Nankai University, Tianjin 300457, China}
\author{Jiafang Li}
\affiliation{Institute of Physics, Chinese Academy of Sciences/Beijing National Laboratory for Condensed Matter Physics, Beijing 100190, China}
\author{Fang Bo}
\affiliation{MOE Key Laboratory of Weak-Light Nonlinear Photonics, TEDA Institute of Applied Physics and School of Physics, Nankai University, Tianjin 300457, China}
\author{Zhong Wang}
\affiliation{Institute for Advanced Study, Tsinghua University, Beijing 100084, China}
\author{Ling Lu}
\email{linglu@iphy.ac.cn. $^\dagger$The first two authors contributed equally.}
\affiliation{Institute of Physics, Chinese Academy of Sciences/Beijing National Laboratory for Condensed Matter Physics, Beijing 100190, China}
\affiliation{Songshan Lake Materials Laboratory, Dongguan, Guangdong 523808, China}

\begin{abstract}
Cavity design is crucial for single-mode semiconductor lasers such as the distributed feedback~(DFB) and vertical-cavity surface-emitting lasers~(VCSEL). By recognizing that both optical resonators feature a single mid-gap mode localized at the topological defect in a one-dimensional~(1D) lattice, we generalize the topological cavity design into 2D using a honeycomb photonic crystal with a vortex Dirac mass --- the analog of Jackiw-Rossi zero modes. 
We theoretically predict and experimentally demonstrate that such a Dirac-vortex cavity can have a tunable mode area across a few orders of magnitudes, arbitrary mode degeneracy, robustly large free-spectral-range, vector-beam output of low divergence, and compatibility with high-index substrates. This topological cavity could enable photonic crystal surface-emitting lasers~(PCSEL) with stabler single-mode operation.
\end{abstract}
\maketitle

Single-mode diode lasers~\cite{chuang2009physics} are the standard light sources for numerous applications, in which the single-modeness relies on the cavity design with subwavelength features.
In long-haul fiber networks, the most widely used DFB laser~\cite{kogelnik1971stimulated}~(Table \ref{tab::dfb}) of a uniform Bragg grating have two competing band-edge modes. Although the mode selection could be done with a certain yield by facet cleaving, a much stabler cavity design is to introduce a quarter-wavelength shift~\cite{haus1976antisymmetric,sekartedjo19841}, so that a single mid-gap mode can lase at the Bragg frequency.
The same 1D mid-gap defect state is also adopted for VCSELs~\cite{chuang2009physics} to select a single longitudinal mode, used in local communications, computer mice, laser printers and face recognitions.
In 2D~\cite{wang1973two,imada1999}, the PCSEL~(Table \ref{tab::dfb}) has recently been commercialized~\cite{HAMAMATSU2018} for its higher power and higher brightness~\cite{matsubara2008gan,hirose2014watt,yoshida2019double}.
However, PCSELs again have at least two high quality-factor~(Q) band-edge modes competing for lasing.
It is obviously important to have a 2D cavity of a single robust mid-gap mode, which has been lacking since the notion of 2D DBFs~\cite{wang1973two}. A stabler lasing mode generally implies higher yield, wider tuning range, narrower linewidth and higher output power. 

In order to design the 2D mid-gap defect cavity, we first recognize that the mid-gap modes of both the phase-shift DFB and VCSEL are in fact topological and are mathematically equivalent to the 1D Jackiw-Rebbi kink state~\cite{jackiw1976} and the Su-Schrieffer-Heeger~(SSH) boundary mode~\cite{su1979solitons}.
This topological view leads us to the Jackiw-Rossi zero mode in 2D~\cite{JACKIW1981}, which we realize in a Dirac photonic crystal with a mass vortex in the silicon-on-insulator~(SOI) platform experimently.

\begin{table}[t!]
\begin{ruledtabular}
\begin{tabular}{>{\centering}m{1.5cm}|m{3.4cm}|m{3.5cm}}
Mode&Bloch band-edge ($\bullet$)
&Topological mid-gap (\textcolor{purple}{$\bullet$})\\\hline
1D&DFB & Phase-shifted DFB \\
&(\textbf{commercialized}) & (\textbf{commercialized})\\
Edge emission&\includegraphics[width=0.19\textwidth]{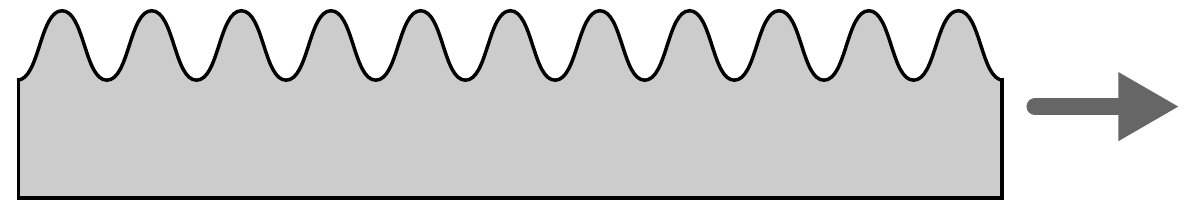}
&\includegraphics[width=0.19\textwidth]{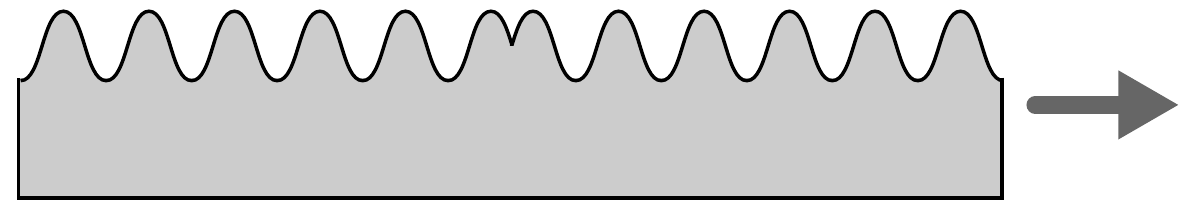}\\
First order feedback&
\begin{tikzpicture}[scale=0.4]
    \draw[black,->] (-2,0) -- (2,0) node[below] {$ka$};
    \draw[black,->] (0,0)node[below] {$\pi$} -- (0,3.5) node[right] {$\omega$};
    \draw[gray,thick] (-2,3.5) parabola bend (0,2.5) (2,3.5);
    \draw[gray,thick] (-2,0.5) parabola bend (0,1.5) (2,0.5);
    \draw [black,fill] (0,2.5) circle [radius=0.15];
    \draw [black,fill] (0,1.5) circle [radius=0.15];
    \draw (2,2)node[black,right,text width=2.1cm]{Two guided modes};
\end{tikzpicture}  
&
\begin{tikzpicture}[scale=0.4]
    \draw[black,->] (-2,0) -- (2,0) node[below] {$ka$};
    \draw[black,->] (0,0)node[below] {$\pi$} -- (0,3.5) node[right] {$\omega$};
    \draw[gray,thick] (-2,3.5) parabola bend (0,2.5) (2,3.5);
    \draw[gray,thick] (-2,0.5) parabola bend (0,1.5) (2,0.5);
    \draw [fill,purple] (0,2) circle [radius=0.15];
    \draw (2,2)node[black,right,text width=2.1cm]{VCSEL\\Jackiw-Rebbi\\SSH};
\end{tikzpicture}  
\\\hline
2D&PCSEL&Dirac-vortex cavity\\
&(\textbf{commercialized}) & (\textbf{this work})\\
{Surface emission}&\includegraphics[width=0.19\textwidth]{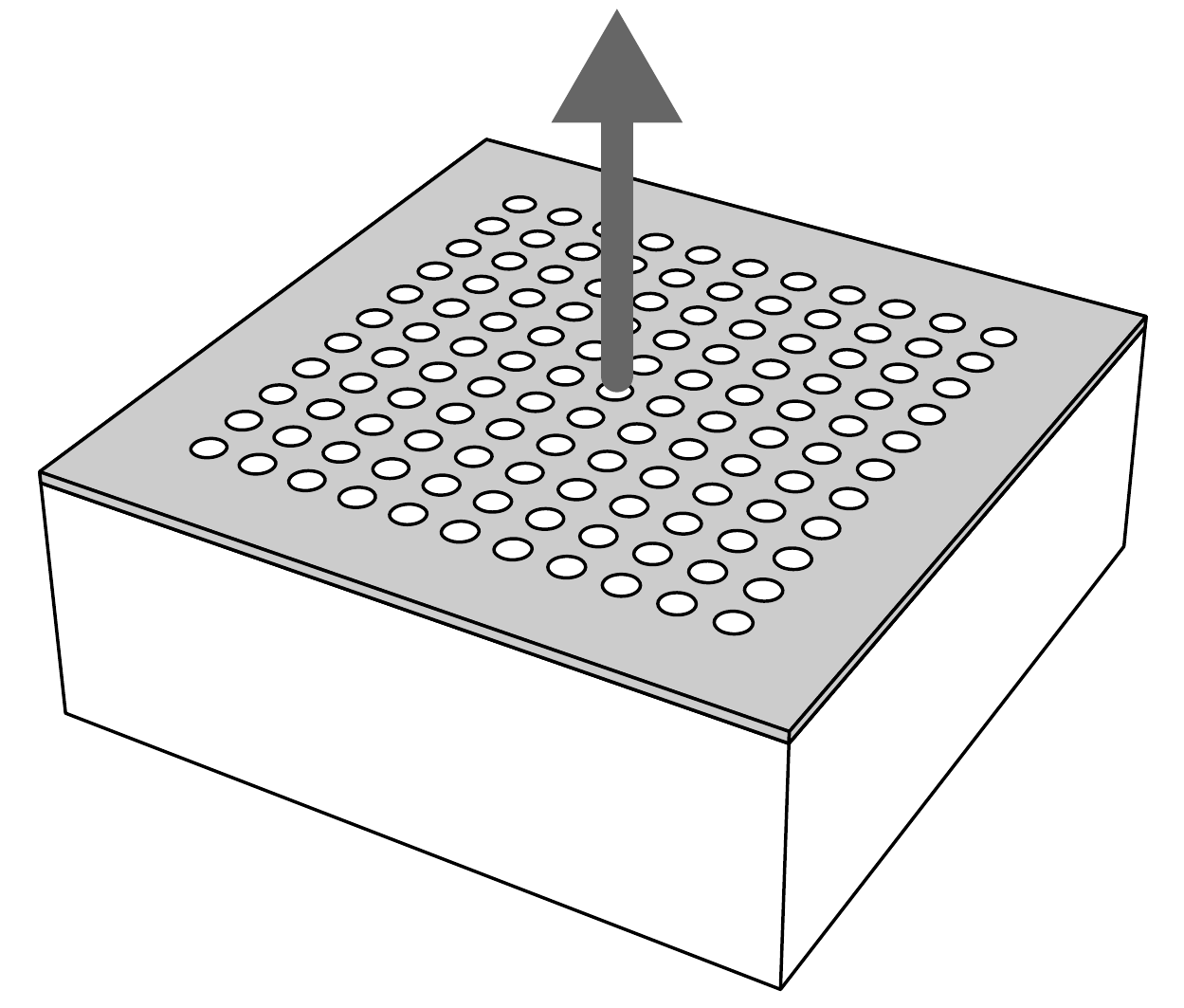}
&\includegraphics[width=0.19\textwidth]{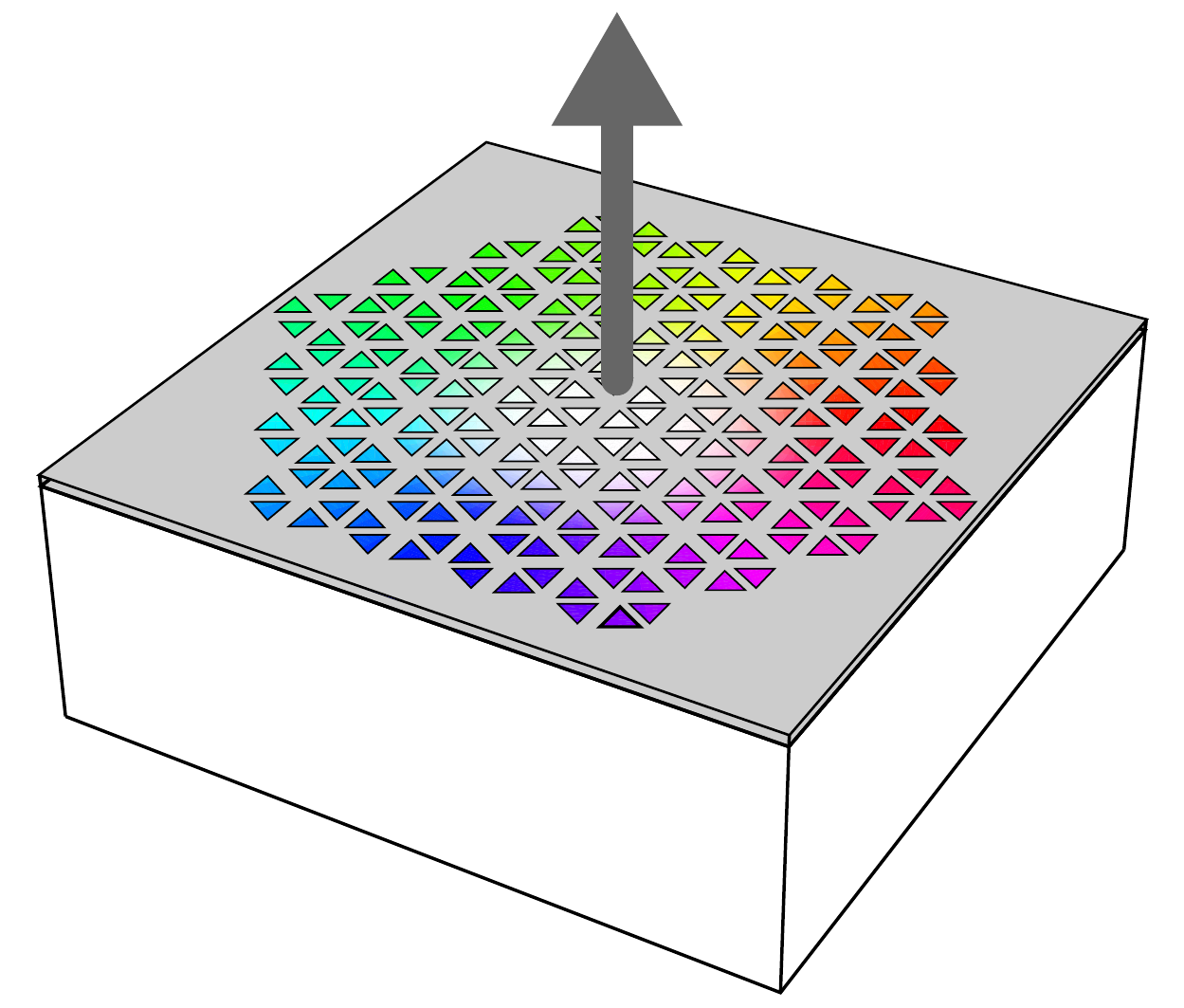}\\
Second order feedback&
\begin{tikzpicture}[scale=0.4]
    \draw[black,->] (-2,0) -- (2,0) node[below] {$ka$};
    \draw[black,->] (0,0)node[below] {$\Gamma$} -- (0,3.5) node[right] {$\omega$};
    \draw[gray,thick] (-2,3.5) parabola bend (0,2.5) (2,3.5);
    \draw[gray,thick] (-1.6,3.5) parabola bend (0,2.5) (1.6,3.5);
    \draw[gray,thick] (-2,0.5) parabola bend (0,1.5) (2,0.5);
    \draw[gray,thick] (-2,1) parabola bend (0,2) (2,1);
    \draw [black,fill] (0,2) circle [radius=0.15];
    \draw [black,fill] (0,1.5) circle [radius=0.15];
    \draw (2,2)node[black,right,text width=2.1cm]{Two high-$Q$ modes};
\end{tikzpicture}  
&
\begin{tikzpicture}[scale=0.4]
    \draw[black,->] (-2,0) -- (2,0) node[below] {$ka$};
    \draw[black,->] (0,0)node[below] {$\Gamma$} -- (0,3.5) node[right] {$\omega$};
    \draw[gray,thick] (-2,3.5) parabola bend (0,2.5) (2,3.5);
    \draw[gray,thick] (-1.6,3.5) parabola bend (0,2.5) (1.6,3.5);
    \draw[gray,thick] (-2,0.5) parabola bend (0,1.5) (2,0.5);
    \draw[gray,thick] (-1.6,0.5) parabola bend (0,1.5) (1.6,0.5);
    \draw [fill,purple] (0,2) circle [radius=0.15];
    \draw (2,2)node[black,right,text width=2.1cm]{Jackiw-Rossi};
\end{tikzpicture}
\\\hline
Advantage&Simple fabrication&Stable operation\\
\end{tabular}
\end{ruledtabular}
\caption{Comparison of the Dirac-vortex cavity and the three types of commercialized semiconductor laser cavities for single-polarization and single-mode operation.
The cavities of uniform lattices, in both 1D DFB and 2D PCSEL, have two band-edge modes of similar thresholds.
The phase-shifted DFB cavity has a single mid-gap mode with the lowest threshold.
Similarly, Dirac-vortex design could stabilize PCSEL. $a$ is the lattice constant.}
\label{tab::dfb}
\end{table}

\begin{figure*}[ht!]
\includegraphics[width=\textwidth]{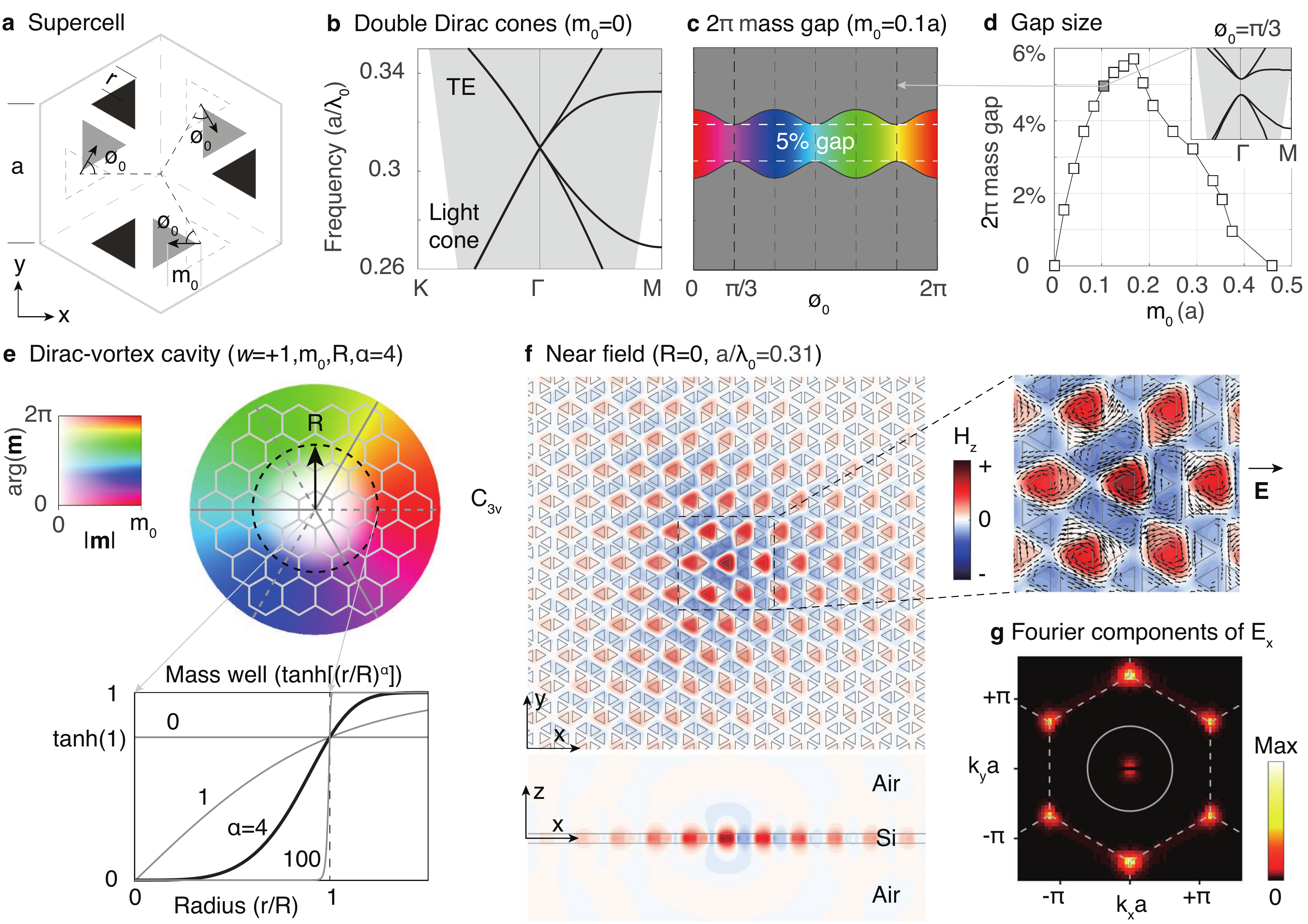}
\caption{Design of the photonic-crystal Dirac-vortex cavity in an air-clad silicon membrane~($0.46a$ thick, $n=3.4$) by 3D simulations for the TE-like modes.
(a) Honeycomb super cell of the generalized Kekul\'e perturbation, where $r=0.32a$.
(b) Double Dirac cone band structure of the unperturbed supercell.
(c) Bandgap opens for $2\pi$ angle of $\phi_0$.
(d) Bandgap size as a function of $m_0$.
(e) Illustration of the Dirac-vortex cavity and the mass-well function.
(f) Near field~($H_z$) of the topological mode with $m_0=0.1a$, $Q$=317, $V=4.0(\lambda_0/n)^3$ and far-field half angle of $4.3^{\circ}$. A central region is magnified with the electric fields plotted.
(g) The magnitudes of the Fourier-transformed $E_x$ fields. The Brillouin zone boundary of the primitive cell and the light cone is outlined.
 }
\label{fig::design}
\end{figure*}

\section{Jackiw-Rossi zero modes}
\label{sec::kp}
The mid-gap modes of the Dirac-vortex cavity is the photonic realization of the zero-mode solutions to the 2D Dirac equations with mass vortices, proposed by Jackiw and Rossi~\cite{JACKIW1981}. 
\begin{equation}
H(\bm{k})=(\sigma_xk_x+\sigma_zk_y)\tau_z+m_1\tau_x+m_2\tau_y+\cancel{m'\sigma_y\tau_z}
\label{eq::Hamiltonian}
\end{equation}
This time-reversal invariant Dirac Hamiltonian in Eq.~\ref{eq::Hamiltonian} contains all five anti-commuting terms, where $\sigma_i$ and $\tau_i$ are Pauli matrices.
As can be seen from the energy eigen-solution $E(\bm{k})=\pm\sqrt{\sum_i(k_i^2+m_i^2)}$, the two momentum terms~($k_i$) in Eq.~\ref{eq::Hamiltonian} form 4-by-4 massless Dirac cones in 2D.
The three mass terms represent three independent mathematical degrees of freedom that can gap the double Dirac cones, acquiring nonzero band curvatures known as the effective masses. If the system has only two mass terms, a vortex solution can form by spatially winding the mass terms in plane. Fortunately, the third mass term $m'$ vanishes when the dispersion spectrum is up-down symmetric with respect to the Dirac frequency. This protecting symmetry is the chiral symmetry $\mathcal{S}=\sigma_y\tau_z$~($\mathcal{S} H \mathcal{S}^{-1} = -H$) whose presence requires $m'=0$. Then the remaining two mass terms form a complex number~[$\mathbf{m}=m_1+jm_2$] that can wind in-plane $w$ times as $\mathbf{m}(\mathbf{r})\propto \textrm{exp}[j w \arg(\mathbf{r})]$, in which  $\mathbf{r}$ is the spatial coordinate and $j^2=-1$. $w$ is the Dirac-mass winding number, the topological invariant of the vortex~\cite{teo2010topological} belonging to the Altland-Zirnbauer symmetry class BDI~($\mathbb{Z}$).
The amplitude and sign of $w$ determine the number and chirality of the mid-gap modes.
We note that in a realistic photonic system at a finite frequency~(instead of zero), the $\mathcal{S}$ is slightly broken and $m'$ is not precisely zero. The resulting Dirac spectrum is not exactly up-down symmetric and the $w$ topological modes are not rigorously degenerate in frequency.

\section{Photonic crystal design}
\label{sec::design}
The design intuition of the Jackiw-Ross modes in realistic systems, from the above analytics, is to start with the double Dirac cones and modulate the lattice to generate the $2\pi$ vortex mass gap for confining the mid-gap modes.
Hou et. al. first suggested that the Dirac-vortex mid-gap modes could be found in a Kekul\'e-textured graphene~\cite{hou2007electron}. Although creating a vortex potential at the atomic level is a tall order, the realization in controlled photonic or phononic lattices~\cite{iadecola2016non,menssen2019photonic,chen2019mechanical,noh2019braiding} has a clear advantage.
In this work, we design the Jackiw-Rossi mid-gap modes in a 2D photonic-crystal silicon membrane of 220~nm thick at 1.55$\mu m$ wavelength.
We first design it with air cladding, then evaluate its performance on substrates.
For efficient computations, all models are up-down symmetric~($z$-mirror), so that the modes can be classified by mirror eigenvalues. In this paper, we focus on the TE-like modes~(transverse-electric, electric field in-plane) that are favored for most applications.

The starting point, in Fig. \ref{fig::design}a, is a hexagon supercell consisting of three honeycomb primitive cells. This supercell folds the two Dirac points from Brillouin-zone boundary~($\pm K$ points below light cone) to the zone center~($\Gamma$ point above light cone), forming a 4-by-4 double Dirac cone dispersion shown in Fig. \ref{fig::design}b. The two honeycomb sub-lattices are colored in black and gray, both representing air-holes in the silicon membrane.
The triangular shape of the air holes, compared to the circular shape, improves the frequency-isolation of the Dirac points~\cite{barik2016,barik2018}. We note that the previous waveguide design~\cite{wu2015scheme} between two deformed honeycomb lattices, by expanding and shrinking, actually corresponds to two discrete phase values of the Dirac mass~(0 and 60$^\circ$).

We apply a generalized Kekul\'e modulation~\cite{hou2007electron} in the supercell to generate the 2$\pi$ vortex mass~(complex mass term with complete $2\pi$ phase) to gap the double Dirac cones. Shown in Fig. \ref{fig::design}a, the three gray sub-lattice~(air holes) in the supercell are shifted from their original positions by the same amplitude of $m_0$ and correlated phase of $\phi_0$.
The key observation from the simulation result, shown in Fig. \ref{fig::design}c, is the persistent gap opening for all 2$\pi$ values of $\phi_0$ with non-zero $m_0$. The gap closes at the vortex center where $m_0=0$.
Due to the symmetry of the supercell, the mass gap in Fig. \ref{fig::design}c has an angular periodicity of $\pi/3$ and the minimal gap size occurs at $\phi_0=\pi/3$.
The gap size as a function of $m_0$ is plotted in Fig. \ref{fig::design}d. The 2$\pi$-mass-gap peaks at 6\% and eventually closes for large $m_0$ because the band at $M$ point drops. 
Since the modulation vector $\mathbf{m}=m_0e^{j\phi_0}$ has the same physical consequence as that of the complex Dirac mass $\mathbf{m}=m_1+jm_2$ in Eq. \ref{eq::Hamiltonian}, we use the same symbol in this paper.

Now that we have a continuous library of supercells with a mass gap for an entire $2\pi$ range of $\phi_0$, the vortex cavity design is a matter of arranging these supercells angularly around a cavity center~($\mathbf{r}_0$), as illustrated in Fig.~\ref{fig::design}e.
Since the original honeycomb lattice~($m_0=0$) have $C_{6v}$ symmetry, the vortex cavity~($m_0\ne0$) can always remain $C_{3v}$ symmetric if a $w$-dependent symmetric vortex center~($\mathbf{r}_0$) is chosen. A highly symmetric design reduces the computation domain and eases the analysis through group theory.

The topological mid-gap mode is plotted in Fig.~\ref{fig::design}f.
The in-plane electric fields form spatial vortices, indicating the vector-beam far fields plotted in Fig.~\ref{fig::2D}. The Fourier components of the mode~{$|FT(E_x)|$} clearly reveals its momentum distribution in relation to the light cone. Once the $K$ points move inside the light cone of the substrate, the Dirac vortex resonance is no longer well defined~(see Sec.~\ref{sec:substrate}).

\section{Cavity parameters}

There is a large degree of freedom in designing the vortex mass $\mathbf{m}(\mathbf{r}-\mathbf{r}_0)$. Without loss of generality, we  choose the form of Eq. \ref{eq::mass-well}.
\begin{equation}
\mathbf{m}(\mathbf{r}-\mathbf{r}_0;w,m_0,R,\alpha)=m_0 \mathrm{tanh}(|\frac{\mathbf{r}-\mathbf{r}_0}{R}|^\alpha) e^{j[\phi_0-w \arg(\mathbf{r}-\mathbf{r}_0)]}
\label{eq::mass-well}
\end{equation}
The mass-well function $\mathrm{tanh}(x^\alpha)|_{x\to+\infty}=+1$ and $\mathrm{tanh}(x^\alpha)|_{x\to0}=x^\alpha$, interpreting from the central zero mass $|\mathbf{m}(\mathbf{r}=\mathbf{r}_0)|=0$ to the boundary maximum mass $|\mathbf{m}(\mathbf{r}\gg\mathbf{r}_0)|=m_0$.
This Dirac-vortex cavity is determined by four parameters~($w,m_0,R,\alpha$) as illustrated in Fig. \ref{fig::design}e.

The first parameter $w$ is the winding number of the vortex.
The magnitude $|w|$ determines the number~(degeneracy) of mid-gap modes and the mode area generally increase with $w$, similar to the topological fiber case~\cite{lu2018topological}.
The sign of $w$ is the mode chirality, determining the field distribution on the sub-lattices~\cite{JACKIW1981}.
The topological mode populates only one of the honeycomb sub-lattice and populates the other sub-lattice when $w$ changes sign.
This can be seen in Fig. \ref{fig::design}f, where both the magnetic~($H_z$) and electric~($E_{x,y}$) fields peak only at the triangles pointing to the left.

The second parameter $m_0 $ is the maximum Dirac mass, the depth of the mass well in Fig. \ref{fig::design}e.
$m_0$ is the maximum shift of the honeycomb sub-lattice in Fig. \ref{fig::design}a, which should be greater than the size of fabrication disorder.
$m_0$ is also the strength of radiative coupling that couples the two (originally guided) Dirac cones into the light cone~(radiation continuum). Therefore the cavity $Q$ increases as $m_0$ decreases. 

The third parameter $R$ is the radius of the vortex, as illustrated in Fig. \ref{fig::design}e.
$R$ should not be mistaken as the size of the whole cavity, outside which the photonic-crystal pattern ends.
We pad at least fifty extra periods outside the vortex radius $R$ to ensure sufficient mode confinement.
We emphasize that $R$ can very different from the size of the confined optical mode. For example, in Fig. \ref{fig::design}f, the mode size is non-zero while the vortex size is ($R=0$). 
Also, the mode size also does not necessarily grow as fast as the vortex size; it depends on $\alpha$.

The fourth parameter $\alpha$ is the shape factor --- a positive exponent that controls the shape of the mass well, plotted in Fig. \ref{fig::design}e.
As a result, $\alpha$ also controls the near field envelope and the radiation pattern of the cavity mode.
We choose $\alpha=4$ in experiments as a balance between radiative coupling and modal size scaling, as discussed in the next section.

\section{Size scaling properties}

\begin{figure}[h]
\includegraphics[width=0.5\textwidth]{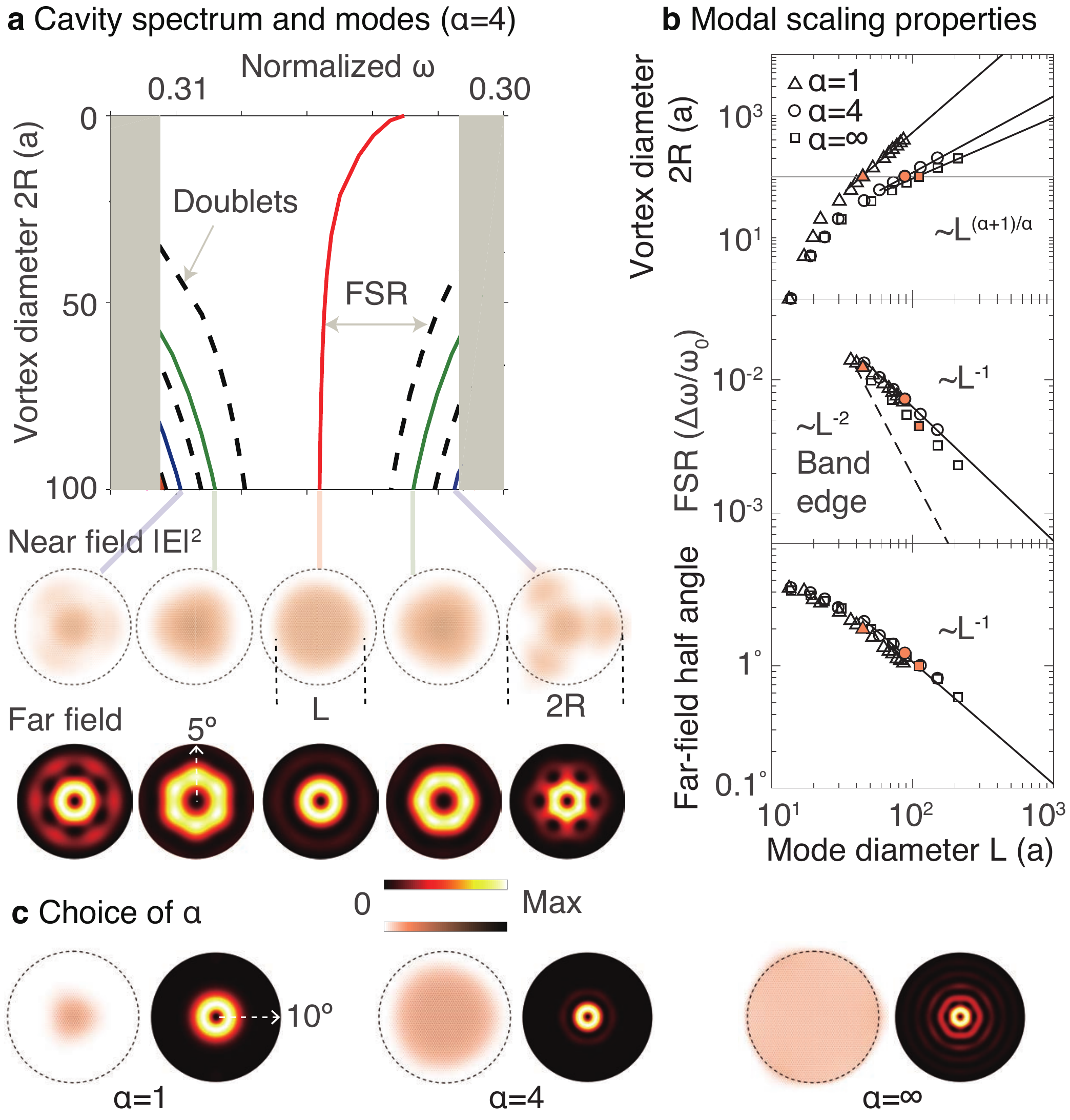}
\caption{Dirac-vortex cavity properties as a function of the vortex size studied using 2D calculations with the effective refractive index of 2.6.
a) The cavity spectrum and modal profiles.
b) 
At the large mode limit,
the mode diameter~($L$) scales as $\propto R^\frac{\alpha}{\alpha+1}$;
the FSR scales as $\propto L^{-1}$;
the far-field angle scales as $\propto L^{-1}$.
When computing $L$, the boundary of the mode is defined at the outer edge where the field intensity drops to $1/e$ of the central maximum intensity.
c) Comparison of the near and far fields for different $\alpha$ with the vortex diameter $2R=100a$.
}
\label{fig::2D}
\end{figure}

For high-power single-mode lasers, one prefers a cavity with broader modal area~(or modal diameter $L$), larger free spectral range~(FSR) and narrower beam divergence. These properties of the Dirac-vortex cavity are thus examined in Fig.~\ref{fig::2D} to see how they scale with the vortex size $R$ using effective 2D simulations. We find that the scaling properties are controlled by the shape factor $\alpha$.

We set $w=+1$ for a single topological mode and choose a large mass gap~($m_0=0.1a$) for large FSRs.
A typical cavity spectrum is shown in Fig.~\ref{fig::2D}a.
For small cavities, the topological mode does not appear exactly at the gap center, due to the lack of chiral symmetry discussed in Sec.~\ref{sec::kp}. 
For large cavities, the topological mode always converges to the Dirac-point frequency, since the central area of the large cavity is approaching the unmodulated Dirac lattice with the original Dirac spectrum.
As $R$ increases, the high-order non-topological cavity modes originate from the continuum of bulk modes above or below the bandgap. These high-order modes have both doublet and singlet states, due to the $C_{3v}$ symmetry. The near fields and far fields of the singlet modes are plotted in Fig.~\ref{fig::2D}a.
The topological mode always has the largest and most uniform mode area, so that it will always experience the largest modal gain given a matching pumping area comparing to the non-topological modes in the cavity.

The modal diameter~($L$) increases with the vortex diameter~($2R$). For large $L$, the scaling is $L\propto R^\frac{\alpha}{\alpha+1}$ shown in Fig.~\ref{fig::2D}b. This is derived from the known result~\cite{JACKIW1981} that the zero mode wavefunction
$\Psi_0(r)$ is determined by the radial integration of the mass function: 
 $|\Psi_0(r)| 
 \propto e^{-\int_0^r |\mathbf{m}(r')|dr'} 
 \propto e^{-\int_0^r (r'/R)^\alpha dr'}
 \propto e^{-\frac{r^{\alpha+1}}{R^\alpha}} $, according the mass definition in Eq.~\ref{eq::mass-well}.
The size of the topological mode grows sub-linearly with $R$ for finite $\alpha$.
Although $\alpha=\infty$ gives the ideal linear size scaling, the cavity is not emitting in the vortex area where $m_0=0$~(meaning zero radiative coupling). The radiation, when $\alpha=\infty$, only takes place at the step boundary, which is not ideal for the power input/output and the far field has multiple fringes as shown in Fig.~\ref{fig::2D}c. Therefore, we choose the shape factor $\alpha=4$ for its narrow far field and near linear scaling of $L\propto R^\frac{4}{5}$.

Dirac-vortex cavity has a robustly large FSR that is essential for the single-mode operation.
As denoted in Fig.~\ref{fig::2D}a, the FSR of the Dirac-vortex cavity is the frequency separation between the mid-gap and the neighboring (doublet) modes. 
It has been pointed out~\cite{chua2014larger} that the FSR of a linear Dirac band edge~($\propto L^{-1}$) is much larger than the FSR of the usual quadratic  band edge~($\propto L^{-2}$), and can be arbitrarily larger in the large mode limit. However, the proposed accidental ``Dirac point" at $\Gamma$ in Ref.~\cite{chua2014larger} is not robust to any system parameters, which means one can never, in reality, fabricate a device and operate at the exact accidental point consistently. 
In contrast, shown in Fig.~\ref{fig::2D}b, our Dirac-vortex cavity has the same $L^{-1}$ advantage for large FSR and this scaling is topologically robust against perturbations to any system parameters!

The far-fields of the singlet modes are vector-beams, as shown in Fig.~\ref{fig::2D}a, obtained by integrating the near-fields using the Rayleigh-Sommerfeld diffraction theory.
Since the polarization-degenerate free-space modes belong to the doublet representation of $C_{3v}$, the singlet cavity modes cannot couple out in the exact vertical direction due to the distinct representations.
If the $C_{3v}$ cavity symmetry is broken, one could convert the donut beam to a single-lobe beam~\cite{miyai2006photonics}.
The beam angle is inversely proportional to the mode diameter in the large mode limit, as plotted in Fig.~\ref{fig::2D}b. The far-field half angle is below 1$^\circ$ once the vortex diameter exceeds 200$a$.

\section{Cavity on substrates}
\label{sec:substrate}

\begin{figure}[b]
\includegraphics[width=0.43\textwidth]{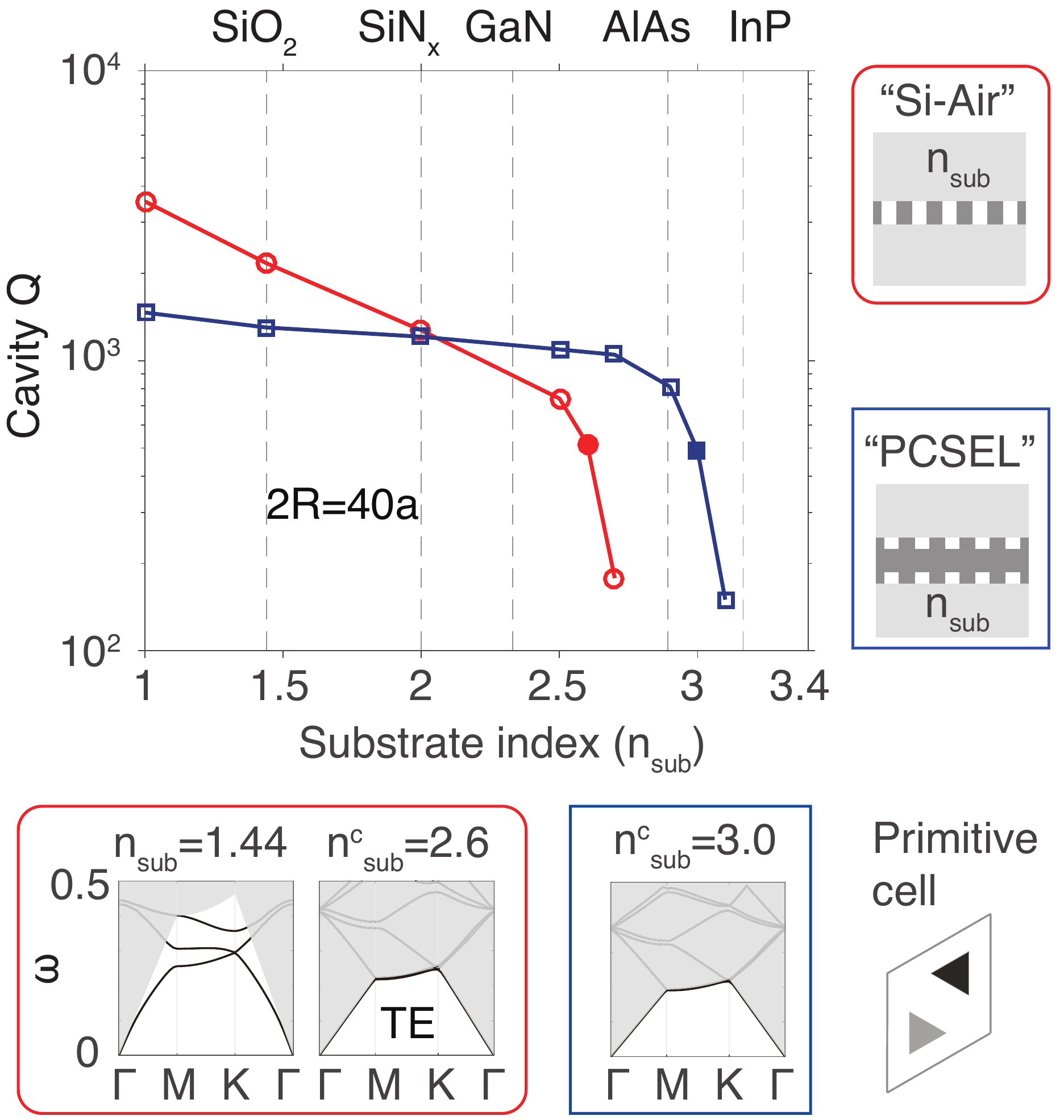}
\caption{Cavity $Q$ as a function of substrate index~($n_{sub}$) studied by 3D FDTD. 
The central photonic-crystal waveguides are made of high-index material~($n=3.4$) and air~($n=1$).
The $Q$ spoils once $n_{sub}>n_{sub}^c$, when the Dirac point enters the light cone in the band structures. The vortex size is $2R=40a$~($\alpha=4$) in this study.
}
\label{fig::substrate}
\end{figure}

\begin{figure*}[ht!]
\includegraphics[width=\textwidth]{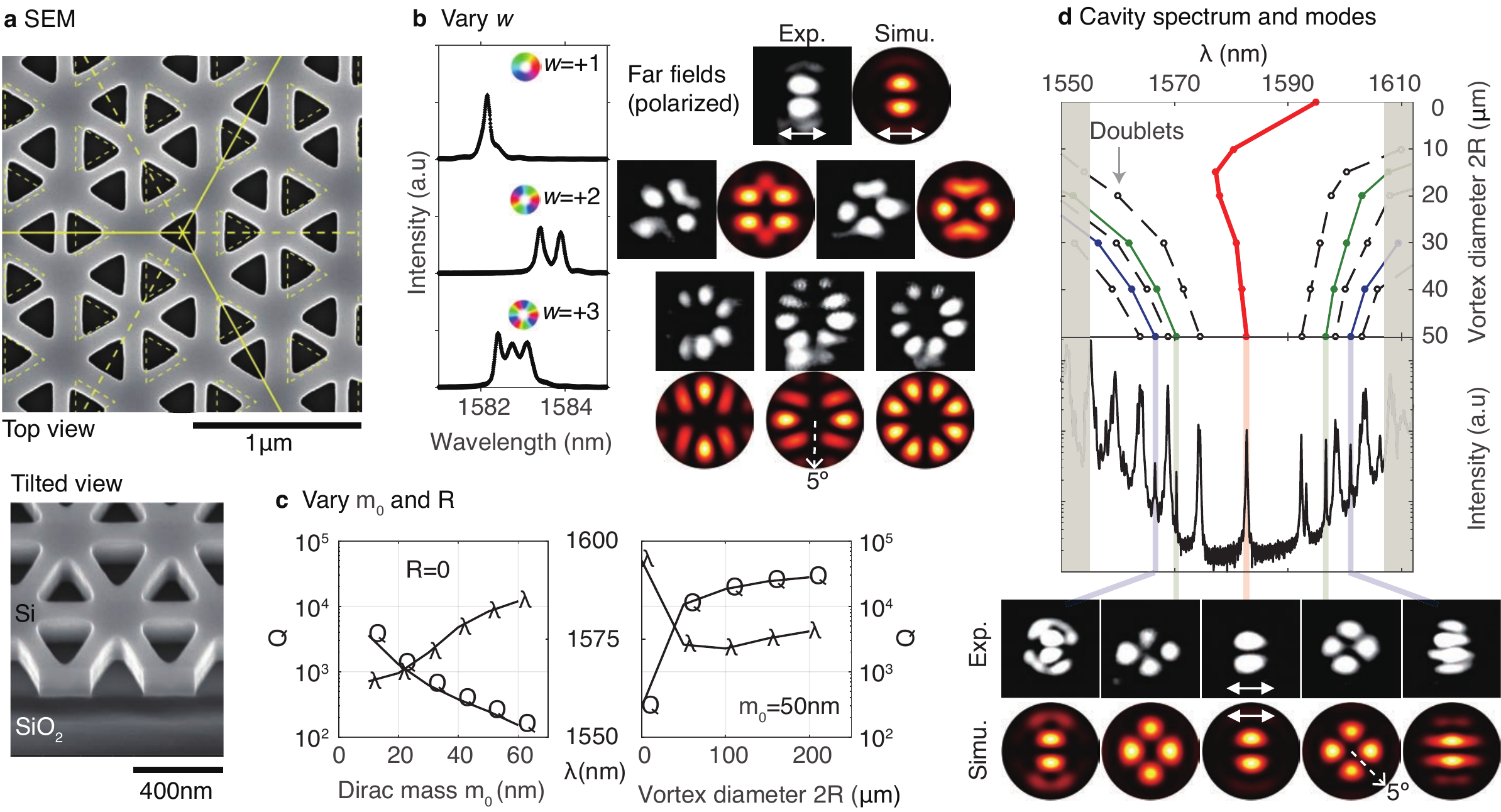}
\caption{Experimental studies of Dirac-vortex cavities with $\alpha=4$.
a) SEM images of a cavity with $R=0\mu$m, $m_0=50nm$ and $w=+1$. The yellow lines illustrate the $C_{3v}$ symmetry and the relative shifts of each air hole.
b) Optical spectra and far fields of cavity modes of different winding numbers with $2R=50\mu m$ and $m_0=50nm$. The $Q$s of the six modes are between $5\times10^3$ and $1\times10^4$.
c) The $Q$s and resonant wavelengths~($\lambda$) of the single-vortex~($w=+1$) cavities measured as a function of Dirac mass $m_0$ and vortex size $R$.
d) The cavity spectrum as a function of vortex size with $w=+1$ and $m_0=50nm$.
The far fields of five singlet modes are imaged and compared with simulations.
}
\label{fig::exp}
\end{figure*}

As a practical device, Dirac-vortex cavities can work on various substrates that dissipate heat, conduct current and provide mechanical support.
In Fig.~\ref{fig::substrate}, we place the cavity on uniform substrates and compute the $Q$ as a function of substrate refractive index~($n_{sub}$) for two different core waveguide configurations: ``Si-Air" and ``PCSEL". 
Both high-index~($n=3.4$) core waveguides are pattern with air. Both types of cavities have a limited vortex size of $2R=40a$ and are top-down symmetric to save computation resources.
Note the cavity $Q$ still increases for larger vortex sizes.

In the ``Si-Air" configuration, we place the silicon-membrane studied in Fig.~\ref{fig::design} on substrates.
The cavity $Q$ gradually decreases~(in power law) with increasing $n_{sub}$ until a critical index value $n^c_{sub}=2.6$ where $Q$ drops exponentially.
This critical point is where the Dirac-point states, in the unperturbed primitive cell , are no longer guided in the core waveguide.
This is shown in the band structures in Fig.~\ref{fig::substrate}, where the Dirac points almost merge into the light cones. This $n^c_{sub}$ value already covers the common substrates such as silica, sapphire and gallium nitride.

In the ``PCSEL" configuration, we aim to further increase $n^c_{sub}$ and evaluate the technological potential of Dirac-vortex PCSELs. The data in Fig.~\ref{fig::substrate} shows $n^c_{sub}=3.0$, which implies the compatibility with the current GaAs/AlGaAs material system used for PCSEL products~\cite{HAMAMATSU2018}. We note that GaAs actually has a higher index of 3.55~(than 3.4 used in our simulation) and our design is not optimized.
Here, the ``PCSEL" waveguide is twice as thick as the ``Si-Air" waveguide with the air-hole pattern through half of its total thickness from top and bottom. (The asymmetric design of moving air holes to one side does not really change the results, but significantly increases the computation time.) This structure is very close to the epitaxy layers of the current PCSEL devices with air-hole maintained regrowth technique~\cite{hirose2014watt,taylor2017optimisation}.

It is worth pointing out that the topological resonance persists even when the Dirac point is not frequency isolated. Actually, when the mode area is large enough, the wavevectors of the mode are too localized (in momentum space) to couple to the other bulk states at the same frequency.
We also note that high cavity $Q$~($\gg10^3$) is undesirable for a high-power laser.
Instead, a good laser cavity emits all its optical loss in the output direction.
The Dirac vortex cavity has the suitable $Q$ and emit only in the vertical direction.
The bottom emission could be easily reflected using a metal coating or Bragg mirrors.

\section{SOI experiments}

We perform the experiments on standard SOI~\cite{hafezi2013imaging,shalaev2018robust,he2019silicon} at telecommunication wavelength to study the spectral and modal properties of the Dirac-vortex cavities with the shape factor $\alpha=4$. 
The scanning-electron-microscope~(SEM) images of a typical device is shown in Fig.~\ref{fig::exp}a. The photonic crystals were patterned in a 220~nm silicon layer by e-beam lithography and dry-etching. The underneath $\rm{SiO_2}$~($n=1.44$) cladding provides mechanical stability. The lattice constant~($a$) is 490~nm.

In Fig.~\ref{fig::exp}b, cavities of different winding number $w=+1,+2,+3$ are measured. Their spectra verify that the number of topological modes equals the winding number. The far fields of all topological modes compare favorably with our simulation results. These radiation patterns are captured after a horizontal polarizer in our cross-polarization setup. Consequently, the number of zero-intensity radial lines equals the topological charges~(in magnitude) of these vector beams. We then focus on the single-mode case $w=+1$ for the rest of the study.

In Fig.~\ref{fig::exp}c, we plot the dependence of $Q$ and wavelength~($\lambda$) on the maximum Dirac mass~($m_0$) and the vortex diameter~($2R$). In both cases, $Q$ increases with the increase of mode area. Because the mode area increases with the decrease of Dirac mass gap and the increase of vortex size.

Shown in Fig.~\ref{fig::exp}d are the cavity spectra as a function of the vortex diameter. Consistent with the numerical results in Fig.~\ref{fig::2D}a, the wavelength of topological mode converges to the Dirac wavelength when the vortex diameter increases to about 30$\mu$m. We also track the high-order modes and a full spectrum is plotted for the cavity of $2R=50\mu m$. The polarized far fields of the singlet modes are imaged and are in agreement with the numerical results.

\section{Discussion}
With the advance of topological photonics~\cite{lu2014topological,khanikaev2017two,ozawa2019topological}, we are able to design a new on-chip  optical microcavity~\cite{vahala2003optical} with separate controls over mode number~($w$), mode area~($R$), radiation coupling~($m_0$) and far-field pattern~($\alpha$), which could outperform other cavities including the topological corner modes~\cite{noh2018topological,ota2019photonic,mittal2019photonic} for topological lasers~\cite{bahari2017nonreciprocal,harari2018topological,bandres2018topological}.
For example, the corner or boundary modes are harder to scale uniformly in area, compared to the vortex cavity.

The Dirac-vortex cavity is the 2D upgrade of the 1D feedback structures in phase-shifted DFB and VCSEL, two widely used industrial semiconductor lasers.
This topological cavity provides a single mid-gap mode with a large modal diameter continuously tunable from a couple of microns towards millimeter scale.
In this work, we have studied the detailed optical properties of this passive cavity.

Next, by changing the lithography pattern, the Dirac-vortex PCSEL could be realized in the same III-V semiconductor platform as that of the current PCSELs~\cite{colombelli2003quantum,HAMAMATSU2018,matsubara2008gan,hirose2014watt,yoshida2019double}. 
The design advantages of a topological PCSEL could be as follows. 
i) It provides a unique single mid-gap mode for lasing.
ii) It has a much larger FSR. Generally, FSR is much larger at the center of the linear Dirac spectrum, where the DOS vanishes, than that at the quadratic band edge where the DOS is a constant in 2D.
iii) The hexagonal lattice generates more uniform in-plane feedback than that of the square lattice in current PCSEL devices. 
In fact, the square lattice was chosen over the hexagonal lattice only to reduce the number of high-Q band edge modes~(2 instead of 4) and suppress multi-mode lasing, which is no longer an issue for the Dirac-vortex cavity.
iv) The cavity design and the above benefits are topologically robust against fabrication errors.
Finally, the employment of the Dirac-vortex cavity in PCSEL devices may lead to stabler operation or even brighter lasers.

\begin{acknowledgements}
We thank Lin Gan and Yong Liang for helpful discussions.
L.L. acknowledges his Ph.D advisor John D. O'Brien~(1969-2017) for his teaching of photonic crystal lasers.
L.L. was supported by the National key R\&D Program of China under Grant No. 2017YFA0303800, 2016YFA0302400 and by NSFC under Project No. 11721404.
Z.W. was supported by NSFC under Grant No. 11674189.
F.B. was supported by NSFC under Grant No. 11734009 and 11674181.
\end{acknowledgements}

\bibliography{references}

\begin{thebibliography}{40}%
\makeatletter
\providecommand \@ifxundefined [1]{%
 \@ifx{#1\undefined}
}%
\providecommand \@ifnum [1]{%
 \ifnum #1\expandafter \@firstoftwo
 \else \expandafter \@secondoftwo
 \fi
}%
\providecommand \@ifx [1]{%
 \ifx #1\expandafter \@firstoftwo
 \else \expandafter \@secondoftwo
 \fi
}%
\providecommand \natexlab [1]{#1}%
\providecommand \enquote  [1]{``#1''}%
\providecommand \bibnamefont  [1]{#1}%
\providecommand \bibfnamefont [1]{#1}%
\providecommand \citenamefont [1]{#1}%
\providecommand \href@noop [0]{\@secondoftwo}%
\providecommand \href [0]{\begingroup \@sanitize@url \@href}%
\providecommand \@href[1]{\@@startlink{#1}\@@href}%
\providecommand \@@href[1]{\endgroup#1\@@endlink}%
\providecommand \@sanitize@url [0]{\catcode `\\12\catcode `\$12\catcode
  `\&12\catcode `\#12\catcode `\^12\catcode `\_12\catcode `\%12\relax}%
\providecommand \@@startlink[1]{}%
\providecommand \@@endlink[0]{}%
\providecommand \url  [0]{\begingroup\@sanitize@url \@url }%
\providecommand \@url [1]{\endgroup\@href {#1}{\urlprefix }}%
\providecommand \urlprefix  [0]{URL }%
\providecommand \Eprint [0]{\href }%
\providecommand \doibase [0]{http://dx.doi.org/}%
\providecommand \selectlanguage [0]{\@gobble}%
\providecommand \bibinfo  [0]{\@secondoftwo}%
\providecommand \bibfield  [0]{\@secondoftwo}%
\providecommand \translation [1]{[#1]}%
\providecommand \BibitemOpen [0]{}%
\providecommand \bibitemStop [0]{}%
\providecommand \bibitemNoStop [0]{.\EOS\space}%
\providecommand \EOS [0]{\spacefactor3000\relax}%
\providecommand \BibitemShut  [1]{\csname bibitem#1\endcsname}%
\let\auto@bib@innerbib\@empty
\bibitem [{\citenamefont {Chuang}(2009)}]{chuang2009physics}%
  \BibitemOpen
  \bibfield  {author} {\bibinfo {author} {\bibfnamefont {Shun~Lien}\
  \bibnamefont {Chuang}},\ }\href@noop {} {\emph {\bibinfo {title} {Physics of
  photonic devices}}}\ (\bibinfo  {publisher} {John Wiley \& Sons},\ \bibinfo
  {year} {2009})\ Chap.~\bibinfo {chapter} {11}\BibitemShut {NoStop}%
\bibitem [{\citenamefont {Kogelnik}\ and\ \citenamefont
  {Shank}(1971)}]{kogelnik1971stimulated}%
  \BibitemOpen
  \bibfield  {author} {\bibinfo {author} {\bibfnamefont {H}~\bibnamefont
  {Kogelnik}}\ and\ \bibinfo {author} {\bibfnamefont {CV}~\bibnamefont
  {Shank}},\ }\bibfield  {title} {\enquote {\bibinfo {title} {Stimulated
  emission in a periodic structure},}\ }\href@noop {} {\bibfield  {journal}
  {\bibinfo  {journal} {Applied Physics Letters}\ }\textbf {\bibinfo {volume}
  {18}},\ \bibinfo {pages} {152--154} (\bibinfo {year} {1971})}\BibitemShut
  {NoStop}%
\bibitem [{\citenamefont {Haus}\ and\ \citenamefont
  {Shank}(1976)}]{haus1976antisymmetric}%
  \BibitemOpen
  \bibfield  {author} {\bibinfo {author} {\bibfnamefont {H}~\bibnamefont
  {Haus}}\ and\ \bibinfo {author} {\bibfnamefont {C}~\bibnamefont {Shank}},\
  }\bibfield  {title} {\enquote {\bibinfo {title} {Antisymmetric taper of
  distributed feedback lasers},}\ }\href@noop {} {\bibfield  {journal}
  {\bibinfo  {journal} {IEEE Journal of Quantum Electronics}\ }\textbf
  {\bibinfo {volume} {12}},\ \bibinfo {pages} {532--539} (\bibinfo {year}
  {1976})}\BibitemShut {NoStop}%
\bibitem [{\citenamefont {Sekartedjo}\ \emph {et~al.}(1984)\citenamefont
  {Sekartedjo}, \citenamefont {Eda}, \citenamefont {Furuya}, \citenamefont
  {Suematsu}, \citenamefont {Koyama},\ and\ \citenamefont
  {Tanbun-Ek}}]{sekartedjo19841}%
  \BibitemOpen
  \bibfield  {author} {\bibinfo {author} {\bibfnamefont {K}~\bibnamefont
  {Sekartedjo}}, \bibinfo {author} {\bibfnamefont {N}~\bibnamefont {Eda}},
  \bibinfo {author} {\bibfnamefont {K}~\bibnamefont {Furuya}}, \bibinfo
  {author} {\bibfnamefont {Y}~\bibnamefont {Suematsu}}, \bibinfo {author}
  {\bibfnamefont {F}~\bibnamefont {Koyama}}, \ and\ \bibinfo {author}
  {\bibfnamefont {T}~\bibnamefont {Tanbun-Ek}},\ }\bibfield  {title} {\enquote
  {\bibinfo {title} {1.5 $\mu$m phase-shifted dfb lasers for single-mode
  operation},}\ }\href@noop {} {\bibfield  {journal} {\bibinfo  {journal}
  {Electronics Letters}\ }\textbf {\bibinfo {volume} {20}},\ \bibinfo {pages}
  {80--81} (\bibinfo {year} {1984})}\BibitemShut {NoStop}%
\bibitem [{\citenamefont {Wang}\ and\ \citenamefont
  {Sheem}(1973)}]{wang1973two}%
  \BibitemOpen
  \bibfield  {author} {\bibinfo {author} {\bibfnamefont {Shyh}\ \bibnamefont
  {Wang}}\ and\ \bibinfo {author} {\bibfnamefont {Sang}\ \bibnamefont
  {Sheem}},\ }\bibfield  {title} {\enquote {\bibinfo {title} {Two-dimensional
  distributed-feedback lasers and their applications},}\ }\href@noop {}
  {\bibfield  {journal} {\bibinfo  {journal} {Applied Physics Letters}\
  }\textbf {\bibinfo {volume} {22}},\ \bibinfo {pages} {460--462} (\bibinfo
  {year} {1973})}\BibitemShut {NoStop}%
\bibitem [{\citenamefont {Imada}\ \emph {et~al.}(1999)\citenamefont {Imada},
  \citenamefont {Noda}, \citenamefont {Chutinan}, \citenamefont {Tokuda},
  \citenamefont {Murata},\ and\ \citenamefont {Sasaki}}]{imada1999}%
  \BibitemOpen
  \bibfield  {author} {\bibinfo {author} {\bibfnamefont {Masahiro}\
  \bibnamefont {Imada}}, \bibinfo {author} {\bibfnamefont {Susumu}\
  \bibnamefont {Noda}}, \bibinfo {author} {\bibfnamefont {Alongkarn}\
  \bibnamefont {Chutinan}}, \bibinfo {author} {\bibfnamefont {Takashi}\
  \bibnamefont {Tokuda}}, \bibinfo {author} {\bibfnamefont {Michio}\
  \bibnamefont {Murata}}, \ and\ \bibinfo {author} {\bibfnamefont {Goro}\
  \bibnamefont {Sasaki}},\ }\bibfield  {title} {\enquote {\bibinfo {title}
  {Coherent two-dimensional lasing action in surface-emitting laser with
  triangular-lattice photonic crystal structure},}\ }\href@noop {} {\bibfield
  {journal} {\bibinfo  {journal} {Applied Physics Letters}\ }\textbf {\bibinfo
  {volume} {75}},\ \bibinfo {pages} {316--318} (\bibinfo {year}
  {1999})}\BibitemShut {NoStop}%
\bibitem [{HAM(2018)}]{HAMAMATSU2018}%
  \BibitemOpen
  \href@noop {} {\emph {\bibinfo {title} {Photonic Crystal Surface Emitting
  Laser Diode L13395-04}}},\ \bibinfo {type} {Tech. Rep.}\ (\bibinfo
  {institution} {HAMAMATSU PHOTONICS K.K.},\ \bibinfo {year}
  {2018})\BibitemShut {NoStop}%
\bibitem [{\citenamefont {Matsubara}\ \emph {et~al.}(2008)\citenamefont
  {Matsubara}, \citenamefont {Yoshimoto}, \citenamefont {Saito}, \citenamefont
  {Jianglin}, \citenamefont {Tanaka},\ and\ \citenamefont
  {Noda}}]{matsubara2008gan}%
  \BibitemOpen
  \bibfield  {author} {\bibinfo {author} {\bibfnamefont {Hideki}\ \bibnamefont
  {Matsubara}}, \bibinfo {author} {\bibfnamefont {Susumu}\ \bibnamefont
  {Yoshimoto}}, \bibinfo {author} {\bibfnamefont {Hirohisa}\ \bibnamefont
  {Saito}}, \bibinfo {author} {\bibfnamefont {Yue}\ \bibnamefont {Jianglin}},
  \bibinfo {author} {\bibfnamefont {Yoshinori}\ \bibnamefont {Tanaka}}, \ and\
  \bibinfo {author} {\bibfnamefont {Susumu}\ \bibnamefont {Noda}},\ }\bibfield
  {title} {\enquote {\bibinfo {title} {Gan photonic-crystal surface-emitting
  laser at blue-violet wavelengths},}\ }\href@noop {} {\bibfield  {journal}
  {\bibinfo  {journal} {Science}\ }\textbf {\bibinfo {volume} {319}},\ \bibinfo
  {pages} {445--447} (\bibinfo {year} {2008})}\BibitemShut {NoStop}%
\bibitem [{\citenamefont {Hirose}\ \emph {et~al.}(2014)\citenamefont {Hirose},
  \citenamefont {Liang}, \citenamefont {Kurosaka}, \citenamefont {Watanabe},
  \citenamefont {Sugiyama},\ and\ \citenamefont {Noda}}]{hirose2014watt}%
  \BibitemOpen
  \bibfield  {author} {\bibinfo {author} {\bibfnamefont {Kazuyoshi}\
  \bibnamefont {Hirose}}, \bibinfo {author} {\bibfnamefont {Yong}\ \bibnamefont
  {Liang}}, \bibinfo {author} {\bibfnamefont {Yoshitaka}\ \bibnamefont
  {Kurosaka}}, \bibinfo {author} {\bibfnamefont {Akiyoshi}\ \bibnamefont
  {Watanabe}}, \bibinfo {author} {\bibfnamefont {Takahiro}\ \bibnamefont
  {Sugiyama}}, \ and\ \bibinfo {author} {\bibfnamefont {Susumu}\ \bibnamefont
  {Noda}},\ }\bibfield  {title} {\enquote {\bibinfo {title} {Watt-class
  high-power, high-beam-quality photonic-crystal lasers},}\ }\href@noop {}
  {\bibfield  {journal} {\bibinfo  {journal} {Nature Photonics}\ }\textbf
  {\bibinfo {volume} {8}},\ \bibinfo {pages} {406} (\bibinfo {year}
  {2014})}\BibitemShut {NoStop}%
\bibitem [{\citenamefont {Yoshida}\ \emph {et~al.}(2019)\citenamefont
  {Yoshida}, \citenamefont {De~Zoysa}, \citenamefont {Ishizaki}, \citenamefont
  {Tanaka}, \citenamefont {Kawasaki}, \citenamefont {Hatsuda}, \citenamefont
  {Song}, \citenamefont {Gelleta},\ and\ \citenamefont
  {Noda}}]{yoshida2019double}%
  \BibitemOpen
  \bibfield  {author} {\bibinfo {author} {\bibfnamefont {Masahiro}\
  \bibnamefont {Yoshida}}, \bibinfo {author} {\bibfnamefont {Menaka}\
  \bibnamefont {De~Zoysa}}, \bibinfo {author} {\bibfnamefont {Kenji}\
  \bibnamefont {Ishizaki}}, \bibinfo {author} {\bibfnamefont {Yoshinori}\
  \bibnamefont {Tanaka}}, \bibinfo {author} {\bibfnamefont {Masato}\
  \bibnamefont {Kawasaki}}, \bibinfo {author} {\bibfnamefont {Ranko}\
  \bibnamefont {Hatsuda}}, \bibinfo {author} {\bibfnamefont {Bongshik}\
  \bibnamefont {Song}}, \bibinfo {author} {\bibfnamefont {John}\ \bibnamefont
  {Gelleta}}, \ and\ \bibinfo {author} {\bibfnamefont {Susumu}\ \bibnamefont
  {Noda}},\ }\bibfield  {title} {\enquote {\bibinfo {title} {Double-lattice
  photonic-crystal resonators enabling high-brightness semiconductor lasers
  with symmetric narrow-divergence beams},}\ }\href@noop {} {\bibfield
  {journal} {\bibinfo  {journal} {Nature Materials}\ }\textbf {\bibinfo
  {volume} {18}},\ \bibinfo {pages} {121} (\bibinfo {year} {2019})}\BibitemShut
  {NoStop}%
\bibitem [{\citenamefont {Jackiw}\ and\ \citenamefont
  {Rebbi}(1976)}]{jackiw1976}%
  \BibitemOpen
  \bibfield  {author} {\bibinfo {author} {\bibfnamefont {R}~\bibnamefont
  {Jackiw}}\ and\ \bibinfo {author} {\bibfnamefont {C}~\bibnamefont {Rebbi}},\
  }\bibfield  {title} {\enquote {\bibinfo {title} {Solitons with fermion number
  1/2},}\ }\href@noop {} {\bibfield  {journal} {\bibinfo  {journal} {Phys. Rev.
  D}\ }\textbf {\bibinfo {volume} {13}},\ \bibinfo {pages} {3398} (\bibinfo
  {year} {1976})}\BibitemShut {NoStop}%
\bibitem [{\citenamefont {Su}\ \emph {et~al.}(1979)\citenamefont {Su},
  \citenamefont {Schrieffer},\ and\ \citenamefont {Heeger}}]{su1979solitons}%
  \BibitemOpen
  \bibfield  {author} {\bibinfo {author} {\bibfnamefont {WP}~\bibnamefont
  {Su}}, \bibinfo {author} {\bibfnamefont {JR}~\bibnamefont {Schrieffer}}, \
  and\ \bibinfo {author} {\bibfnamefont {Ao~J}\ \bibnamefont {Heeger}},\
  }\bibfield  {title} {\enquote {\bibinfo {title} {Solitons in
  polyacetylene},}\ }\href@noop {} {\bibfield  {journal} {\bibinfo  {journal}
  {Physical Review Letters}\ }\textbf {\bibinfo {volume} {42}},\ \bibinfo
  {pages} {1698} (\bibinfo {year} {1979})}\BibitemShut {NoStop}%
\bibitem [{\citenamefont {Jackiw}\ and\ \citenamefont
  {Rossi}(1981)}]{JACKIW1981}%
  \BibitemOpen
  \bibfield  {author} {\bibinfo {author} {\bibfnamefont {R.}~\bibnamefont
  {Jackiw}}\ and\ \bibinfo {author} {\bibfnamefont {P.}~\bibnamefont {Rossi}},\
  }\bibfield  {title} {\enquote {\bibinfo {title} {Zero modes of the
  vortex-fermion system},}\ }\href@noop {} {\bibfield  {journal} {\bibinfo
  {journal} {Nuclear Physics B}\ }\textbf {\bibinfo {volume} {190}},\ \bibinfo
  {pages} {681 -- 691} (\bibinfo {year} {1981})}\BibitemShut {NoStop}%
\bibitem [{\citenamefont {Teo}\ and\ \citenamefont
  {Kane}(2010)}]{teo2010topological}%
  \BibitemOpen
  \bibfield  {author} {\bibinfo {author} {\bibfnamefont {Jeffrey~CY}\
  \bibnamefont {Teo}}\ and\ \bibinfo {author} {\bibfnamefont {Charles~L}\
  \bibnamefont {Kane}},\ }\bibfield  {title} {\enquote {\bibinfo {title}
  {Topological defects and gapless modes in insulators and superconductors},}\
  }\href@noop {} {\bibfield  {journal} {\bibinfo  {journal} {Physical Review
  B}\ }\textbf {\bibinfo {volume} {82}},\ \bibinfo {pages} {115120} (\bibinfo
  {year} {2010})}\BibitemShut {NoStop}%
\bibitem [{\citenamefont {Hou}\ \emph {et~al.}(2007)\citenamefont {Hou},
  \citenamefont {Chamon},\ and\ \citenamefont {Mudry}}]{hou2007electron}%
  \BibitemOpen
  \bibfield  {author} {\bibinfo {author} {\bibfnamefont {Chang-Yu}\
  \bibnamefont {Hou}}, \bibinfo {author} {\bibfnamefont {Claudio}\ \bibnamefont
  {Chamon}}, \ and\ \bibinfo {author} {\bibfnamefont {Christopher}\
  \bibnamefont {Mudry}},\ }\bibfield  {title} {\enquote {\bibinfo {title}
  {Electron fractionalization in two-dimensional graphenelike structures},}\
  }\href@noop {} {\bibfield  {journal} {\bibinfo  {journal} {Physical Review
  Letters}\ }\textbf {\bibinfo {volume} {98}},\ \bibinfo {pages} {186809}
  (\bibinfo {year} {2007})}\BibitemShut {NoStop}%
\bibitem [{\citenamefont {Iadecola}\ \emph {et~al.}(2016)\citenamefont
  {Iadecola}, \citenamefont {Schuster},\ and\ \citenamefont
  {Chamon}}]{iadecola2016non}%
  \BibitemOpen
  \bibfield  {author} {\bibinfo {author} {\bibfnamefont {Thomas}\ \bibnamefont
  {Iadecola}}, \bibinfo {author} {\bibfnamefont {Thomas}\ \bibnamefont
  {Schuster}}, \ and\ \bibinfo {author} {\bibfnamefont {Claudio}\ \bibnamefont
  {Chamon}},\ }\bibfield  {title} {\enquote {\bibinfo {title} {Non-abelian
  braiding of light},}\ }\href@noop {} {\bibfield  {journal} {\bibinfo
  {journal} {Physical Review Letters}\ }\textbf {\bibinfo {volume} {117}},\
  \bibinfo {pages} {073901} (\bibinfo {year} {2016})}\BibitemShut {NoStop}%
\bibitem [{\citenamefont {Menssen}\ \emph {et~al.}(2019)\citenamefont
  {Menssen}, \citenamefont {Guan}, \citenamefont {Felce}, \citenamefont
  {Booth},\ and\ \citenamefont {Walmsley}}]{menssen2019photonic}%
  \BibitemOpen
  \bibfield  {author} {\bibinfo {author} {\bibfnamefont {Adrian~J}\
  \bibnamefont {Menssen}}, \bibinfo {author} {\bibfnamefont {Jun}\ \bibnamefont
  {Guan}}, \bibinfo {author} {\bibfnamefont {David}\ \bibnamefont {Felce}},
  \bibinfo {author} {\bibfnamefont {Martin~J}\ \bibnamefont {Booth}}, \ and\
  \bibinfo {author} {\bibfnamefont {Ian~A}\ \bibnamefont {Walmsley}},\
  }\bibfield  {title} {\enquote {\bibinfo {title} {A photonic majorana bound
  state},}\ }\href@noop {} {\bibfield  {journal} {\bibinfo  {journal} {arXiv
  preprint arXiv:1901.04439}\ } (\bibinfo {year} {2019})}\BibitemShut {NoStop}%
\bibitem [{\citenamefont {Chen}\ \emph {et~al.}(2019)\citenamefont {Chen},
  \citenamefont {Lera}, \citenamefont {Chaunsali}, \citenamefont {Torrent},
  \citenamefont {Alvarez}, \citenamefont {Yang}, \citenamefont {San-Jose},\
  and\ \citenamefont {Christensen}}]{chen2019mechanical}%
  \BibitemOpen
  \bibfield  {author} {\bibinfo {author} {\bibfnamefont {Chun-Wei}\
  \bibnamefont {Chen}}, \bibinfo {author} {\bibfnamefont {Natalia}\
  \bibnamefont {Lera}}, \bibinfo {author} {\bibfnamefont {Rajesh}\ \bibnamefont
  {Chaunsali}}, \bibinfo {author} {\bibfnamefont {Daniel}\ \bibnamefont
  {Torrent}}, \bibinfo {author} {\bibfnamefont {Jose~Vicente}\ \bibnamefont
  {Alvarez}}, \bibinfo {author} {\bibfnamefont {Jinkyu}\ \bibnamefont {Yang}},
  \bibinfo {author} {\bibfnamefont {Pablo}\ \bibnamefont {San-Jose}}, \ and\
  \bibinfo {author} {\bibfnamefont {Johan}\ \bibnamefont {Christensen}},\
  }\bibfield  {title} {\enquote {\bibinfo {title} {Mechanical analogue of a
  majorana bound state},}\ }\href@noop {} {\bibfield  {journal} {\bibinfo
  {journal} {arXiv preprint arXiv:1905.03510}\ } (\bibinfo {year}
  {2019})}\BibitemShut {NoStop}%
\bibitem [{\citenamefont {Noh}\ \emph {et~al.}(2019)\citenamefont {Noh},
  \citenamefont {Schuster}, \citenamefont {Iadecola}, \citenamefont {Huang},
  \citenamefont {Wang}, \citenamefont {Chen}, \citenamefont {Chamon},\ and\
  \citenamefont {Rechtsman}}]{noh2019braiding}%
  \BibitemOpen
  \bibfield  {author} {\bibinfo {author} {\bibfnamefont {Jiho}\ \bibnamefont
  {Noh}}, \bibinfo {author} {\bibfnamefont {Thomas}\ \bibnamefont {Schuster}},
  \bibinfo {author} {\bibfnamefont {Thomas}\ \bibnamefont {Iadecola}}, \bibinfo
  {author} {\bibfnamefont {Sheng}\ \bibnamefont {Huang}}, \bibinfo {author}
  {\bibfnamefont {Mohan}\ \bibnamefont {Wang}}, \bibinfo {author}
  {\bibfnamefont {Kevin~P}\ \bibnamefont {Chen}}, \bibinfo {author}
  {\bibfnamefont {Claudio}\ \bibnamefont {Chamon}}, \ and\ \bibinfo {author}
  {\bibfnamefont {Mikael~C}\ \bibnamefont {Rechtsman}},\ }\bibfield  {title}
  {\enquote {\bibinfo {title} {Braiding photonic topological zero modes},}\
  }\href@noop {} {\bibfield  {journal} {\bibinfo  {journal} {arXiv preprint
  arXiv:1907.03208}\ } (\bibinfo {year} {2019})}\BibitemShut {NoStop}%
\bibitem [{\citenamefont {Barik}\ \emph {et~al.}(2016)\citenamefont {Barik},
  \citenamefont {Miyake}, \citenamefont {DeGottardi}, \citenamefont {Waks},\
  and\ \citenamefont {Hafezi}}]{barik2016}%
  \BibitemOpen
  \bibfield  {author} {\bibinfo {author} {\bibfnamefont {Sabyasachi}\
  \bibnamefont {Barik}}, \bibinfo {author} {\bibfnamefont {Hirokazu}\
  \bibnamefont {Miyake}}, \bibinfo {author} {\bibfnamefont {Wade}\ \bibnamefont
  {DeGottardi}}, \bibinfo {author} {\bibfnamefont {Edo}\ \bibnamefont {Waks}},
  \ and\ \bibinfo {author} {\bibfnamefont {Mohammad}\ \bibnamefont {Hafezi}},\
  }\bibfield  {title} {\enquote {\bibinfo {title} {Two-dimensionally confined
  topological edge states in photonic crystals},}\ }\href@noop {} {\bibfield
  {journal} {\bibinfo  {journal} {New Journal of Physics}\ }\textbf {\bibinfo
  {volume} {18}},\ \bibinfo {pages} {113013} (\bibinfo {year}
  {2016})}\BibitemShut {NoStop}%
\bibitem [{\citenamefont {Barik}\ \emph {et~al.}(2018)\citenamefont {Barik},
  \citenamefont {Karasahin}, \citenamefont {Flower}, \citenamefont {Cai},
  \citenamefont {Miyake}, \citenamefont {DeGottardi}, \citenamefont {Hafezi},\
  and\ \citenamefont {Waks}}]{barik2018}%
  \BibitemOpen
  \bibfield  {author} {\bibinfo {author} {\bibfnamefont {Sabyasachi}\
  \bibnamefont {Barik}}, \bibinfo {author} {\bibfnamefont {Aziz}\ \bibnamefont
  {Karasahin}}, \bibinfo {author} {\bibfnamefont {Christopher}\ \bibnamefont
  {Flower}}, \bibinfo {author} {\bibfnamefont {Tao}\ \bibnamefont {Cai}},
  \bibinfo {author} {\bibfnamefont {Hirokazu}\ \bibnamefont {Miyake}}, \bibinfo
  {author} {\bibfnamefont {Wade}\ \bibnamefont {DeGottardi}}, \bibinfo {author}
  {\bibfnamefont {Mohammad}\ \bibnamefont {Hafezi}}, \ and\ \bibinfo {author}
  {\bibfnamefont {Edo}\ \bibnamefont {Waks}},\ }\bibfield  {title} {\enquote
  {\bibinfo {title} {A topological quantum optics interface},}\ }\href@noop {}
  {\bibfield  {journal} {\bibinfo  {journal} {Science}\ }\textbf {\bibinfo
  {volume} {359}},\ \bibinfo {pages} {666--668} (\bibinfo {year}
  {2018})}\BibitemShut {NoStop}%
\bibitem [{\citenamefont {Wu}\ and\ \citenamefont {Hu}(2015)}]{wu2015scheme}%
  \BibitemOpen
  \bibfield  {author} {\bibinfo {author} {\bibfnamefont {Long-Hua}\
  \bibnamefont {Wu}}\ and\ \bibinfo {author} {\bibfnamefont {Xiao}\
  \bibnamefont {Hu}},\ }\bibfield  {title} {\enquote {\bibinfo {title} {Scheme
  for achieving a topological photonic crystal by using dielectric material},}\
  }\href@noop {} {\bibfield  {journal} {\bibinfo  {journal} {Physical Review
  Letters}\ }\textbf {\bibinfo {volume} {114}},\ \bibinfo {pages} {223901}
  (\bibinfo {year} {2015})}\BibitemShut {NoStop}%
\bibitem [{\citenamefont {Lu}\ \emph {et~al.}(2018)\citenamefont {Lu},
  \citenamefont {Gao},\ and\ \citenamefont {Wang}}]{lu2018topological}%
  \BibitemOpen
  \bibfield  {author} {\bibinfo {author} {\bibfnamefont {Ling}\ \bibnamefont
  {Lu}}, \bibinfo {author} {\bibfnamefont {Haozhe}\ \bibnamefont {Gao}}, \ and\
  \bibinfo {author} {\bibfnamefont {Zhong}\ \bibnamefont {Wang}},\ }\bibfield
  {title} {\enquote {\bibinfo {title} {Topological one-way fiber of second
  chern number},}\ }\href@noop {} {\bibfield  {journal} {\bibinfo  {journal}
  {Nature Communications}\ }\textbf {\bibinfo {volume} {9}},\ \bibinfo {pages}
  {5384} (\bibinfo {year} {2018})}\BibitemShut {NoStop}%
\bibitem [{\citenamefont {Chua}\ \emph {et~al.}(2014)\citenamefont {Chua},
  \citenamefont {Lu}, \citenamefont {Bravo-Abad}, \citenamefont
  {Joannopoulos},\ and\ \citenamefont {Solja{\v{c}}i{\'c}}}]{chua2014larger}%
  \BibitemOpen
  \bibfield  {author} {\bibinfo {author} {\bibfnamefont {Song-Liang}\
  \bibnamefont {Chua}}, \bibinfo {author} {\bibfnamefont {Ling}\ \bibnamefont
  {Lu}}, \bibinfo {author} {\bibfnamefont {Jorge}\ \bibnamefont {Bravo-Abad}},
  \bibinfo {author} {\bibfnamefont {John~D}\ \bibnamefont {Joannopoulos}}, \
  and\ \bibinfo {author} {\bibfnamefont {Marin}\ \bibnamefont
  {Solja{\v{c}}i{\'c}}},\ }\bibfield  {title} {\enquote {\bibinfo {title}
  {Larger-area single-mode photonic crystal surface-emitting lasers enabled by
  an accidental dirac point},}\ }\href@noop {} {\bibfield  {journal} {\bibinfo
  {journal} {Optics Letters}\ }\textbf {\bibinfo {volume} {39}},\ \bibinfo
  {pages} {2072--2075} (\bibinfo {year} {2014})}\BibitemShut {NoStop}%
\bibitem [{\citenamefont {Miyai}\ \emph {et~al.}(2006)\citenamefont {Miyai},
  \citenamefont {Sakai}, \citenamefont {Okano}, \citenamefont {Kunishi},
  \citenamefont {Ohnishi},\ and\ \citenamefont {Noda}}]{miyai2006photonics}%
  \BibitemOpen
  \bibfield  {author} {\bibinfo {author} {\bibfnamefont {Eiji}\ \bibnamefont
  {Miyai}}, \bibinfo {author} {\bibfnamefont {Kyosuke}\ \bibnamefont {Sakai}},
  \bibinfo {author} {\bibfnamefont {Takayuki}\ \bibnamefont {Okano}}, \bibinfo
  {author} {\bibfnamefont {Wataru}\ \bibnamefont {Kunishi}}, \bibinfo {author}
  {\bibfnamefont {Dai}\ \bibnamefont {Ohnishi}}, \ and\ \bibinfo {author}
  {\bibfnamefont {Susumu}\ \bibnamefont {Noda}},\ }\bibfield  {title} {\enquote
  {\bibinfo {title} {Lasers producing tailored beams},}\ }\href@noop {}
  {\bibfield  {journal} {\bibinfo  {journal} {Nature}\ }\textbf {\bibinfo
  {volume} {441}},\ \bibinfo {pages} {946} (\bibinfo {year}
  {2006})}\BibitemShut {NoStop}%
\bibitem [{\citenamefont {Taylor}\ \emph {et~al.}(2017)\citenamefont {Taylor},
  \citenamefont {Ivanov}, \citenamefont {Li}, \citenamefont {Childs},\ and\
  \citenamefont {Hogg}}]{taylor2017optimisation}%
  \BibitemOpen
  \bibfield  {author} {\bibinfo {author} {\bibfnamefont {RJE}\ \bibnamefont
  {Taylor}}, \bibinfo {author} {\bibfnamefont {P}~\bibnamefont {Ivanov}},
  \bibinfo {author} {\bibfnamefont {G}~\bibnamefont {Li}}, \bibinfo {author}
  {\bibfnamefont {DTD}\ \bibnamefont {Childs}}, \ and\ \bibinfo {author}
  {\bibfnamefont {RA}~\bibnamefont {Hogg}},\ }\bibfield  {title} {\enquote
  {\bibinfo {title} {Optimisation of photonic crystal coupling through
  waveguide design},}\ }\href@noop {} {\bibfield  {journal} {\bibinfo
  {journal} {Optical and Quantum Electronics}\ }\textbf {\bibinfo {volume}
  {49}},\ \bibinfo {pages} {47} (\bibinfo {year} {2017})}\BibitemShut {NoStop}%
\bibitem [{\citenamefont {Hafezi}\ \emph {et~al.}(2013)\citenamefont {Hafezi},
  \citenamefont {Mittal}, \citenamefont {Fan}, \citenamefont {Migdall},\ and\
  \citenamefont {Taylor}}]{hafezi2013imaging}%
  \BibitemOpen
  \bibfield  {author} {\bibinfo {author} {\bibfnamefont {Mohammad}\
  \bibnamefont {Hafezi}}, \bibinfo {author} {\bibfnamefont {S}~\bibnamefont
  {Mittal}}, \bibinfo {author} {\bibfnamefont {J}~\bibnamefont {Fan}}, \bibinfo
  {author} {\bibfnamefont {A}~\bibnamefont {Migdall}}, \ and\ \bibinfo {author}
  {\bibfnamefont {JM}~\bibnamefont {Taylor}},\ }\bibfield  {title} {\enquote
  {\bibinfo {title} {Imaging topological edge states in silicon photonics},}\
  }\href@noop {} {\bibfield  {journal} {\bibinfo  {journal} {Nature Photonics}\
  }\textbf {\bibinfo {volume} {7}},\ \bibinfo {pages} {1001} (\bibinfo {year}
  {2013})}\BibitemShut {NoStop}%
\bibitem [{\citenamefont {Shalaev}\ \emph {et~al.}(2018)\citenamefont
  {Shalaev}, \citenamefont {Walasik}, \citenamefont {Tsukernik}, \citenamefont
  {Xu},\ and\ \citenamefont {Litchinitser}}]{shalaev2018robust}%
  \BibitemOpen
  \bibfield  {author} {\bibinfo {author} {\bibfnamefont {Mikhail~I}\
  \bibnamefont {Shalaev}}, \bibinfo {author} {\bibfnamefont {Wiktor}\
  \bibnamefont {Walasik}}, \bibinfo {author} {\bibfnamefont {Alexander}\
  \bibnamefont {Tsukernik}}, \bibinfo {author} {\bibfnamefont {Yun}\
  \bibnamefont {Xu}}, \ and\ \bibinfo {author} {\bibfnamefont {Natalia~M}\
  \bibnamefont {Litchinitser}},\ }\bibfield  {title} {\enquote {\bibinfo
  {title} {Robust topologically protected transport in photonic crystals at
  telecommunication wavelengths},}\ }\href@noop {} {\bibfield  {journal}
  {\bibinfo  {journal} {Nature nanotechnology}\ }\textbf {\bibinfo {volume}
  {14}},\ \bibinfo {pages} {31} (\bibinfo {year} {2018})}\BibitemShut {NoStop}%
\bibitem [{\citenamefont {He}\ \emph {et~al.}(2019)\citenamefont {He},
  \citenamefont {Liang}, \citenamefont {Yuan}, \citenamefont {Qiu},
  \citenamefont {Chen}, \citenamefont {Zhao},\ and\ \citenamefont
  {Dong}}]{he2019silicon}%
  \BibitemOpen
  \bibfield  {author} {\bibinfo {author} {\bibfnamefont {Xin-Tao}\ \bibnamefont
  {He}}, \bibinfo {author} {\bibfnamefont {En-Tao}\ \bibnamefont {Liang}},
  \bibinfo {author} {\bibfnamefont {Jia-Jun}\ \bibnamefont {Yuan}}, \bibinfo
  {author} {\bibfnamefont {Hao-Yang}\ \bibnamefont {Qiu}}, \bibinfo {author}
  {\bibfnamefont {Xiao-Dong}\ \bibnamefont {Chen}}, \bibinfo {author}
  {\bibfnamefont {Fu-Li}\ \bibnamefont {Zhao}}, \ and\ \bibinfo {author}
  {\bibfnamefont {Jian-Wen}\ \bibnamefont {Dong}},\ }\bibfield  {title}
  {\enquote {\bibinfo {title} {A silicon-on-insulator slab for topological
  valley transport},}\ }\href@noop {} {\bibfield  {journal} {\bibinfo
  {journal} {Nature Communications}\ }\textbf {\bibinfo {volume} {10}},\
  \bibinfo {pages} {872} (\bibinfo {year} {2019})}\BibitemShut {NoStop}%
\bibitem [{\citenamefont {Lu}\ \emph {et~al.}(2014)\citenamefont {Lu},
  \citenamefont {Joannopoulos},\ and\ \citenamefont
  {Solja{\v{c}}i{\'c}}}]{lu2014topological}%
  \BibitemOpen
  \bibfield  {author} {\bibinfo {author} {\bibfnamefont {Ling}\ \bibnamefont
  {Lu}}, \bibinfo {author} {\bibfnamefont {John~D}\ \bibnamefont
  {Joannopoulos}}, \ and\ \bibinfo {author} {\bibfnamefont {Marin}\
  \bibnamefont {Solja{\v{c}}i{\'c}}},\ }\bibfield  {title} {\enquote {\bibinfo
  {title} {Topological photonics},}\ }\href@noop {} {\bibfield  {journal}
  {\bibinfo  {journal} {Nature Photonics}\ }\textbf {\bibinfo {volume} {8}},\
  \bibinfo {pages} {821--829} (\bibinfo {year} {2014})}\BibitemShut {NoStop}%
\bibitem [{\citenamefont {Khanikaev}\ and\ \citenamefont
  {Shvets}(2017)}]{khanikaev2017two}%
  \BibitemOpen
  \bibfield  {author} {\bibinfo {author} {\bibfnamefont {Alexander~B}\
  \bibnamefont {Khanikaev}}\ and\ \bibinfo {author} {\bibfnamefont {Gennady}\
  \bibnamefont {Shvets}},\ }\bibfield  {title} {\enquote {\bibinfo {title}
  {Two-dimensional topological photonics},}\ }\href@noop {} {\bibfield
  {journal} {\bibinfo  {journal} {Nature Photonics}\ }\textbf {\bibinfo
  {volume} {11}},\ \bibinfo {pages} {763} (\bibinfo {year} {2017})}\BibitemShut
  {NoStop}%
\bibitem [{\citenamefont {Ozawa}\ \emph {et~al.}(2019)\citenamefont {Ozawa},
  \citenamefont {Price}, \citenamefont {Amo}, \citenamefont {Goldman},
  \citenamefont {Hafezi}, \citenamefont {Lu}, \citenamefont {Rechtsman},
  \citenamefont {Schuster}, \citenamefont {Simon}, \citenamefont {Zilberberg}
  \emph {et~al.}}]{ozawa2019topological}%
  \BibitemOpen
  \bibfield  {author} {\bibinfo {author} {\bibfnamefont {Tomoki}\ \bibnamefont
  {Ozawa}}, \bibinfo {author} {\bibfnamefont {Hannah~M}\ \bibnamefont {Price}},
  \bibinfo {author} {\bibfnamefont {Alberto}\ \bibnamefont {Amo}}, \bibinfo
  {author} {\bibfnamefont {Nathan}\ \bibnamefont {Goldman}}, \bibinfo {author}
  {\bibfnamefont {Mohammad}\ \bibnamefont {Hafezi}}, \bibinfo {author}
  {\bibfnamefont {Ling}\ \bibnamefont {Lu}}, \bibinfo {author} {\bibfnamefont
  {Mikael~C}\ \bibnamefont {Rechtsman}}, \bibinfo {author} {\bibfnamefont
  {David}\ \bibnamefont {Schuster}}, \bibinfo {author} {\bibfnamefont
  {Jonathan}\ \bibnamefont {Simon}}, \bibinfo {author} {\bibfnamefont {Oded}\
  \bibnamefont {Zilberberg}},  \emph {et~al.},\ }\bibfield  {title} {\enquote
  {\bibinfo {title} {Topological photonics},}\ }\href@noop {} {\bibfield
  {journal} {\bibinfo  {journal} {Reviews of Modern Physics}\ }\textbf
  {\bibinfo {volume} {91}},\ \bibinfo {pages} {015006} (\bibinfo {year}
  {2019})}\BibitemShut {NoStop}%
\bibitem [{\citenamefont {Vahala}(2003)}]{vahala2003optical}%
  \BibitemOpen
  \bibfield  {author} {\bibinfo {author} {\bibfnamefont {Kerry~J}\ \bibnamefont
  {Vahala}},\ }\bibfield  {title} {\enquote {\bibinfo {title} {Optical
  microcavities},}\ }\href@noop {} {\bibfield  {journal} {\bibinfo  {journal}
  {Nature}\ }\textbf {\bibinfo {volume} {424}},\ \bibinfo {pages} {839}
  (\bibinfo {year} {2003})}\BibitemShut {NoStop}%
\bibitem [{\citenamefont {Noh}\ \emph {et~al.}(2018)\citenamefont {Noh},
  \citenamefont {Benalcazar}, \citenamefont {Huang}, \citenamefont {Collins},
  \citenamefont {Chen}, \citenamefont {Hughes},\ and\ \citenamefont
  {Rechtsman}}]{noh2018topological}%
  \BibitemOpen
  \bibfield  {author} {\bibinfo {author} {\bibfnamefont {Jiho}\ \bibnamefont
  {Noh}}, \bibinfo {author} {\bibfnamefont {Wladimir~A}\ \bibnamefont
  {Benalcazar}}, \bibinfo {author} {\bibfnamefont {Sheng}\ \bibnamefont
  {Huang}}, \bibinfo {author} {\bibfnamefont {Matthew~J}\ \bibnamefont
  {Collins}}, \bibinfo {author} {\bibfnamefont {Kevin~P}\ \bibnamefont {Chen}},
  \bibinfo {author} {\bibfnamefont {Taylor~L}\ \bibnamefont {Hughes}}, \ and\
  \bibinfo {author} {\bibfnamefont {Mikael~C}\ \bibnamefont {Rechtsman}},\
  }\bibfield  {title} {\enquote {\bibinfo {title} {Topological protection of
  photonic mid-gap defect modes},}\ }\href@noop {} {\bibfield  {journal}
  {\bibinfo  {journal} {Nature Photonics}\ }\textbf {\bibinfo {volume} {12}},\
  \bibinfo {pages} {408} (\bibinfo {year} {2018})}\BibitemShut {NoStop}%
\bibitem [{\citenamefont {Ota}\ \emph {et~al.}(2019)\citenamefont {Ota},
  \citenamefont {Liu}, \citenamefont {Katsumi}, \citenamefont {Watanabe},
  \citenamefont {Wakabayashi}, \citenamefont {Arakawa},\ and\ \citenamefont
  {Iwamoto}}]{ota2019photonic}%
  \BibitemOpen
  \bibfield  {author} {\bibinfo {author} {\bibfnamefont {Yasutomo}\
  \bibnamefont {Ota}}, \bibinfo {author} {\bibfnamefont {Feng}\ \bibnamefont
  {Liu}}, \bibinfo {author} {\bibfnamefont {Ryota}\ \bibnamefont {Katsumi}},
  \bibinfo {author} {\bibfnamefont {Katsuyuki}\ \bibnamefont {Watanabe}},
  \bibinfo {author} {\bibfnamefont {Katsunori}\ \bibnamefont {Wakabayashi}},
  \bibinfo {author} {\bibfnamefont {Yasuhiko}\ \bibnamefont {Arakawa}}, \ and\
  \bibinfo {author} {\bibfnamefont {Satoshi}\ \bibnamefont {Iwamoto}},\
  }\bibfield  {title} {\enquote {\bibinfo {title} {Photonic crystal nanocavity
  based on a topological corner state},}\ }\href@noop {} {\bibfield  {journal}
  {\bibinfo  {journal} {Optica}\ }\textbf {\bibinfo {volume} {6}},\ \bibinfo
  {pages} {786--789} (\bibinfo {year} {2019})}\BibitemShut {NoStop}%
\bibitem [{\citenamefont {Mittal}\ \emph {et~al.}(2019)\citenamefont {Mittal},
  \citenamefont {Orre}, \citenamefont {Zhu}, \citenamefont {Gorlach},
  \citenamefont {Poddubny},\ and\ \citenamefont {Hafezi}}]{mittal2019photonic}%
  \BibitemOpen
  \bibfield  {author} {\bibinfo {author} {\bibfnamefont {Sunil}\ \bibnamefont
  {Mittal}}, \bibinfo {author} {\bibfnamefont {Venkata~Vikram}\ \bibnamefont
  {Orre}}, \bibinfo {author} {\bibfnamefont {Guanyu}\ \bibnamefont {Zhu}},
  \bibinfo {author} {\bibfnamefont {Maxim~A}\ \bibnamefont {Gorlach}}, \bibinfo
  {author} {\bibfnamefont {Alexander}\ \bibnamefont {Poddubny}}, \ and\
  \bibinfo {author} {\bibfnamefont {Mohammad}\ \bibnamefont {Hafezi}},\
  }\bibfield  {title} {\enquote {\bibinfo {title} {Photonic quadrupole
  topological phases},}\ }\href@noop {} {\bibfield  {journal} {\bibinfo
  {journal} {Nature Photonics}\ }\textbf {\bibinfo {volume} {13}},\ \bibinfo
  {pages} {692} (\bibinfo {year} {2019})}\BibitemShut {NoStop}%
\bibitem [{\citenamefont {Bahari}\ \emph {et~al.}(2017)\citenamefont {Bahari},
  \citenamefont {Ndao}, \citenamefont {Vallini}, \citenamefont {El~Amili},
  \citenamefont {Fainman},\ and\ \citenamefont
  {Kant{\'e}}}]{bahari2017nonreciprocal}%
  \BibitemOpen
  \bibfield  {author} {\bibinfo {author} {\bibfnamefont {Babak}\ \bibnamefont
  {Bahari}}, \bibinfo {author} {\bibfnamefont {Abdoulaye}\ \bibnamefont
  {Ndao}}, \bibinfo {author} {\bibfnamefont {Felipe}\ \bibnamefont {Vallini}},
  \bibinfo {author} {\bibfnamefont {Abdelkrim}\ \bibnamefont {El~Amili}},
  \bibinfo {author} {\bibfnamefont {Yeshaiahu}\ \bibnamefont {Fainman}}, \ and\
  \bibinfo {author} {\bibfnamefont {Boubacar}\ \bibnamefont {Kant{\'e}}},\
  }\bibfield  {title} {\enquote {\bibinfo {title} {Nonreciprocal lasing in
  topological cavities of arbitrary geometries},}\ }\href@noop {} {\bibfield
  {journal} {\bibinfo  {journal} {Science}\ }\textbf {\bibinfo {volume}
  {358}},\ \bibinfo {pages} {636--640} (\bibinfo {year} {2017})}\BibitemShut
  {NoStop}%
\bibitem [{\citenamefont {Harari}\ \emph {et~al.}(2018)\citenamefont {Harari},
  \citenamefont {Bandres}, \citenamefont {Lumer}, \citenamefont {Rechtsman},
  \citenamefont {Chong}, \citenamefont {Khajavikhan}, \citenamefont
  {Christodoulides},\ and\ \citenamefont {Segev}}]{harari2018topological}%
  \BibitemOpen
  \bibfield  {author} {\bibinfo {author} {\bibfnamefont {Gal}\ \bibnamefont
  {Harari}}, \bibinfo {author} {\bibfnamefont {Miguel~A}\ \bibnamefont
  {Bandres}}, \bibinfo {author} {\bibfnamefont {Yaakov}\ \bibnamefont {Lumer}},
  \bibinfo {author} {\bibfnamefont {Mikael~C}\ \bibnamefont {Rechtsman}},
  \bibinfo {author} {\bibfnamefont {Yi~Dong}\ \bibnamefont {Chong}}, \bibinfo
  {author} {\bibfnamefont {Mercedeh}\ \bibnamefont {Khajavikhan}}, \bibinfo
  {author} {\bibfnamefont {Demetrios~N}\ \bibnamefont {Christodoulides}}, \
  and\ \bibinfo {author} {\bibfnamefont {Mordechai}\ \bibnamefont {Segev}},\
  }\bibfield  {title} {\enquote {\bibinfo {title} {Topological insulator laser:
  theory},}\ }\href@noop {} {\bibfield  {journal} {\bibinfo  {journal}
  {Science}\ }\textbf {\bibinfo {volume} {359}},\ \bibinfo {pages} {eaar4003}
  (\bibinfo {year} {2018})}\BibitemShut {NoStop}%
\bibitem [{\citenamefont {Bandres}\ \emph {et~al.}(2018)\citenamefont
  {Bandres}, \citenamefont {Wittek}, \citenamefont {Harari}, \citenamefont
  {Parto}, \citenamefont {Ren}, \citenamefont {Segev}, \citenamefont
  {Christodoulides},\ and\ \citenamefont
  {Khajavikhan}}]{bandres2018topological}%
  \BibitemOpen
  \bibfield  {author} {\bibinfo {author} {\bibfnamefont {Miguel~A}\
  \bibnamefont {Bandres}}, \bibinfo {author} {\bibfnamefont {Steffen}\
  \bibnamefont {Wittek}}, \bibinfo {author} {\bibfnamefont {Gal}\ \bibnamefont
  {Harari}}, \bibinfo {author} {\bibfnamefont {Midya}\ \bibnamefont {Parto}},
  \bibinfo {author} {\bibfnamefont {Jinhan}\ \bibnamefont {Ren}}, \bibinfo
  {author} {\bibfnamefont {Mordechai}\ \bibnamefont {Segev}}, \bibinfo {author}
  {\bibfnamefont {Demetrios~N}\ \bibnamefont {Christodoulides}}, \ and\
  \bibinfo {author} {\bibfnamefont {Mercedeh}\ \bibnamefont {Khajavikhan}},\
  }\bibfield  {title} {\enquote {\bibinfo {title} {Topological insulator laser:
  Experiments},}\ }\href@noop {} {\bibfield  {journal} {\bibinfo  {journal}
  {Science}\ }\textbf {\bibinfo {volume} {359}},\ \bibinfo {pages} {eaar4005}
  (\bibinfo {year} {2018})}\BibitemShut {NoStop}%
\bibitem [{\citenamefont {Colombelli}\ \emph {et~al.}(2003)\citenamefont
  {Colombelli}, \citenamefont {Srinivasan}, \citenamefont {Troccoli},
  \citenamefont {Painter}, \citenamefont {Gmachl}, \citenamefont {Tennant},
  \citenamefont {Sergent}, \citenamefont {Sivco}, \citenamefont {Cho},\ and\
  \citenamefont {Capasso}}]{colombelli2003quantum}%
  \BibitemOpen
  \bibfield  {author} {\bibinfo {author} {\bibfnamefont {Raffaele}\
  \bibnamefont {Colombelli}}, \bibinfo {author} {\bibfnamefont {Kartik}\
  \bibnamefont {Srinivasan}}, \bibinfo {author} {\bibfnamefont {Mariano}\
  \bibnamefont {Troccoli}}, \bibinfo {author} {\bibfnamefont {Oskar}\
  \bibnamefont {Painter}}, \bibinfo {author} {\bibfnamefont {Claire~F}\
  \bibnamefont {Gmachl}}, \bibinfo {author} {\bibfnamefont {Donald~M}\
  \bibnamefont {Tennant}}, \bibinfo {author} {\bibfnamefont {A~Michael}\
  \bibnamefont {Sergent}}, \bibinfo {author} {\bibfnamefont {Deborah~L}\
  \bibnamefont {Sivco}}, \bibinfo {author} {\bibfnamefont {Alfred~Y}\
  \bibnamefont {Cho}}, \ and\ \bibinfo {author} {\bibfnamefont {Federico}\
  \bibnamefont {Capasso}},\ }\bibfield  {title} {\enquote {\bibinfo {title}
  {Quantum cascade surface-emitting photonic crystal laser},}\ }\href@noop {}
  {\bibfield  {journal} {\bibinfo  {journal} {Science}\ }\textbf {\bibinfo
  {volume} {302}},\ \bibinfo {pages} {1374--1377} (\bibinfo {year}
  {2003})}\BibitemShut {NoStop}%
\end{thebibliography}%

\end{document}